\documentclass[english]{article}
\usepackage{mathptmx}
\usepackage[T1]{fontenc}
\usepackage[utf8]{inputenc}
\usepackage[a4paper]{geometry}
\usepackage{color}
\usepackage{babel}
\usepackage{array}
\usepackage{float}
\usepackage{units}
\usepackage{bm}
\usepackage{amssymb}
\usepackage{amsmath}
\usepackage{graphicx}
\usepackage{wasysym}
\usepackage{gensymb}
\usepackage{setspace}
\usepackage{textcomp}
\usepackage[backend=biber, style=nature, sorting=none, isbn=false, url=false, eprint=false, maxbibnames=15]{biblatex}
\usepackage[unicode=true,
bookmarks=true,bookmarksnumbered=false,bookmarksopen=false,
 breaklinks=false,pdfborder={0 0 0},pdfborderstyle={},backref=false,colorlinks=true]{hyperref}
\hypersetup{linkcolor=black, citecolor=black, urlcolor=black, filecolor=blue}
\usepackage{textgreek}
\usepackage{csquotes}
\usepackage[font=small,labelfont=bf]{caption}

\AtEveryBibitem{\clearfield{language}}

\begin{document}

\begin{center}
    {\Large 
        \textbf{High-Yield Assembly of Plasmon-Coupled Nanodiamonds \textit{via} DNA Origami for Tailored Emission}
    }
\end{center}

\begin{center}
    {\large{}
        Niklas Hansen$^{1,3}$, 
        Jakub \v{C}op\'{a}k$^{2,4}$, 
        Marek Kindermann$^{2}$,
        David Roesel$^{1}$, 
        Federica Scollo$^{1}$, 
        Ilko Bald$^{1,6}$
        Petr Cigler$^{2*}$, 
        Vladim\'{i}ra Petr\'{a}kov\'{a}$^{1, 5*}$}
\end{center}

\begin{center}
    {\footnotesize $^{1}$J. Heyrovsk\'{y} Institute of Physical Chemistry, Czech Academy of Sciences, Dolej\v{s}kova 3, 182 23 Prague, Czechia}\\
    {\footnotesize $^{2}$Institute of Chemistry and Biochemistry, Czech Academy of Sciences, Flemingovo nam\v{e}st\'{i} 542/2, 160 00 Prague, Czechia}\\
    {\footnotesize $^{3}$Department of Physical Chemistry, Faculty of Chemical Engineering, University of Chemistry and Technology, Technick\'{a} 5, \newline166 28 Prague 6, Czechia}\\
    {\footnotesize $^{4}$Department of Physical and Macromolecular Chemistry, Faculty of Science, Charles University, Hlavova 2030/8, \newline128 40 Prague 2, Czechia}\\
    {\footnotesize $^{5}$ Faculty of Biomedical Engineering, Czech Technical University, Nam\v{e}st\'{i} S\'{i}tn\'{a} 3105, 27201 Kladno 2, Czechia}\\
    {\footnotesize $^{6}$Institut f\"{u}r Chemie, Universit\"{a}t Potsdam, Karl-Liebknecht-Stra{\ss}e 24-25, 14476 Potsdam, Germany}\\
\end{center}

\begin{center}
    $^{*}$Correspondence to: vladimira.petrakova@jh-inst.cas.cz, petr.cigler@uochb.cas.cz
\end{center}

\begin{abstract}
Controlling the spatial arrangement of optically active elements is crucial for the advancement of engineered photonic systems. Color centers in nanodiamond offer unique advantages for quantum sensing and information processing; however, their integration into complex optical architectures is limited by challenges in precise and reproducible positioning, as well as efficient coupling. DNA origami provides an elegant solution, as demonstrated by recent studies showcasing nanoscale positioning of fluorescent nanodiamonds and plasmonic gold nanoparticles. Here, we present a scalable and robust method for covalently functionalizing nanodiamonds with DNA, enabling high-yield, spatially controlled assembly of diamond and gold nanoparticles onto DNA origami. By precisely controlling the interparticle spacing, we reveal distance-dependent modulation of NV center photoluminescence with a 10-fold increase in the fastest decay pathway at short interparticle distances. Our findings indicate selective plasmon-driven effects and interplay between radiative and non-radiative processes. This work overcomes key limitations in current nanodiamond assembly strategies and provides insights into engineering NV photoluminescence by plasmonic coupling that advance toward quantum photonic and sensing applications. 
\end{abstract}

\vfill{}

\newpage{}

\section{Introduction}

Many photonic phenomena rely on the precise spatial arrangement of optically active elements at the nanoscale. Controlled positioning is essential in both natural and engineered systems, enabling efficient energy transfer and light manipulation. In nature, photosynthetic light-harvesting complexes achieve remarkable energy transfer efficiencies through precise molecular organization \cite{mirkovic_light_2017, scholes_using_2017, ernst_photoinduced_2025}. Similarly, engineered photonic systems, such as metamaterials, nanophotonic circuits, and solid-state quantum systems \cite{solntsev_metasurfaces_2021, yang_nanofabrication_2025,mittal_topological_2018, mueller_deep_2020}, rely on nanoscale structuring to achieve tailored optical responses, including engineered quantum entanglement.

One intriguing optical quantum system is the nitrogen-vacancy (NV) center in diamond, whose optical and spin properties are highly sensitive to its close environment. NV centers are among the systems that have exhibited spontaneous superradiance at room temperature, as demonstrated in nanodiamonds containing a high density of NV centers \cite{bradac_room-temperature_2017}. Their long spin coherence times, even at room temperature, make them attractive for quantum sensing and information processing applications 
\cite{herbschleb_ultra-long_2019, gaebel_room-temperature_2006, bar-gill_solid-state_2013}. Furthermore, the emission of NV centers can be tuned \textit{via} surface chemical functionalization, which modulates their charge states and alters their spectral characteristics \cite{hauf_chemical_2011, petrakova_luminescence_2012, petrakova_charge-sensitive_2015, petrakova_imaging_2016}. This high degree of environmental sensitivity enables their applications in nanoscale sensing, in the detection of magnetic and electric fields, temperature variations, and molecular interactions \cite{rovny_nanoscale_2022, rovny_nanoscale_2024, maze_nanoscale_2008, dolde_electric-field_2011, maletinsky_robust_2012, neumann_high-precision_2013, schirhagl_nitrogen-vacancy_2014, hsiao_fluorescent_2016, zhang_toward_2021}.

Coupling diamond color centers to other light-modulating elements, such as plasmonic nanoparticles (NPs), offers further opportunities to tailor their emission properties and enhance their integration into photonic devices \cite{choy_enhanced_2011, bulu_plasmonic_2011, hapuarachchi_plasmonically_2024, ding_purcell-enhanced_2025}. The general focus is on enhancing photoluminescence (PL) of NV centers, which results in increased sensitivity \cite{barry_sensitivity_2020, taylor_high-sensitivity_2008}. Plasmonic structures modify the radiative dynamics of PL defect centers, increasing their spontaneous emission rates and directing emitted photons more efficiently. Examples include enhanced single-photon emission rates \cite{schietinger_plasmon-enhanced_2009}, ultrafast radiative decay \cite{boyce_plasmonic_2024, kolesov_waveparticle_2009}, and polarization engineering \textit{via} plasmonic nanoantennas \cite{kan_metasurfaceenabled_2020, andersen_ultrabright_2017} and nanogrove structures \cite{kumar_excitation_2016}, showing efficient coupling \cite{barth_controlled_2010, yanagimoto_purcell_2021} and excitations \cite{gur_dna-assembled_2018}. Additionally, plasmonic coupling can mediate coherent energy transfer, enabling the controlled routing of optical signals within nanoscale circuits \cite{roller_hotspot-mediated_2017}, offering an extremely promising line of applications for quantum computing and sensing when combined with NV centers. 

A critical limitation in expanding the applications of plasmonically coupled NV centers are difficulties in producing spatially precise and reproducible binding between the centers and plasmonic nanoparticles in sufficient quantities. Many of the achievements relied on fabrication techniques such as random positioning, single-particle manipulation or non-specific binding, which lack precise control over the assembly design or yield sparse structures. Top-down lithography techniques offer an alternative \cite{shegai_bimetallic_2011,mack_decoupling_2017}, but they suffer from limited spatial control in the regime of nanometers, creating nanoscale inhomogeneities caused by the deposition techniques that may significantly alter the strong electromagnetic near-fields and lead to the creation of random hotspots, modifying the plasmonic characteristics. This poses challenges for reproducibility and limits practical applications in constructed devices.

DNA nanotechnology offers an elegant alternative to solve the spatial arrangement, taking advantage of the programmable nature of the spatial configuration of DNA
\cite{rothemund_folding_2006}. This technique enables the use of chemically synthesized plasmonic nanoparticles that are monodisperse and of crystalline quality, making their plasmonic properties uniform and tunable \cite{kuzyk_dna-based_2012,kuzyk_reconfigurable_2014, schreiber_hierarchical_2014, li_colloidal_2022}. DNA origami-based assemblies achieve sub-nanometer precision in positioning of nanoparticles, fluorophores, or other molecules based on the sequence-specific folding of the DNA strands in the desired origami structure. The most common applications are found in decorating origami nanostructures with plasmonic NPs like gold (AuNPs) or silver (AgNP) nanoparticles for fluorescence enhancement \cite{vietz_functionalizing_2016}, biosensing \cite{dass_dna_2021, huang_dna_2018}, and spectroscopy \cite{thacker_dna_2014}. Many different architectures have been demonstrated with a variety of different kinds of nanoparticles, as well as their ability to form metamolecular structures with distinct optical responses \cite{lan_bifacial_2013, liu_dnaorigamibased_2017, lee_dna_2018, chen_nanoscale_2022}. 

The potential of DNA origami for precisely positioning NV center-containing nanodiamonds (NDs) has been demonstrated through advances in recent studies. Zhang et al. \cite{zhang_dna-based_2015} successfully realized nanoscale positioning of fluorescent NDs on DNA origami, using a bioconjugation strategy based on biotinylated PEG-labeled biopolymers to functionalize the NDs, confirming their optical properties remained unaffected by the coating. However, the study also highlighted key limitations, including the need for more appropriate coatings to improve coupling efficiency and greater positioning precision to explore coherent interactions between NV centers. Another study \cite{gur_dna-assembled_2018} demonstrated DNA origami-assembled AuNP chains with sub-2\,nm interparticle spacings, achieving low propagation losses that improved energy transfer efficiency in plasmonic waveguides. The work underscored the challenge of precisely integrating NV centers into plasmonic circuits with controlled spacing to fully harness their quantum optical properties. These achievements demonstrate the vast potential of precisely arranged NV centers for spin-based quantum technologies and plasmon-enhanced sensing, highlighting the need to overcome the practical challenges in assembly to fully take advantage of their capabilities. 
\\
Here, we address these challenges by developing a covalent DNA functionalization strategy that ensures robust binding to DNA origami. We further demonstrate the capabilities of this approach by incorporating plasmonic nanoparticles into the assembly. Through systematic optimization, we establish a DNA-mediated assembly process for nanodiamond-gold nanoparticle hybrids, achieving yields exceeding 50\,\% in most cases. By controlling the separation of gold and diamond nanoparticles, we demonstrate distance-dependent modulation of photoluminescence of NV centers that shows efficient coupling between the NV excited electronic states and localized surface plasmons, leading to over 10-fold enhancement of the fastest decay pathway. Our results bring insight into the interplay between radiative and non-radiative processes in plasmon-coupled NV centers, advancing their potential in quantum photonics and nanoscale sensing.

\section{Results \& Discussion}
\subsection{Concept \& Design}
\begin{figure}[ht]
    \centering
    \includegraphics[width=1\linewidth]{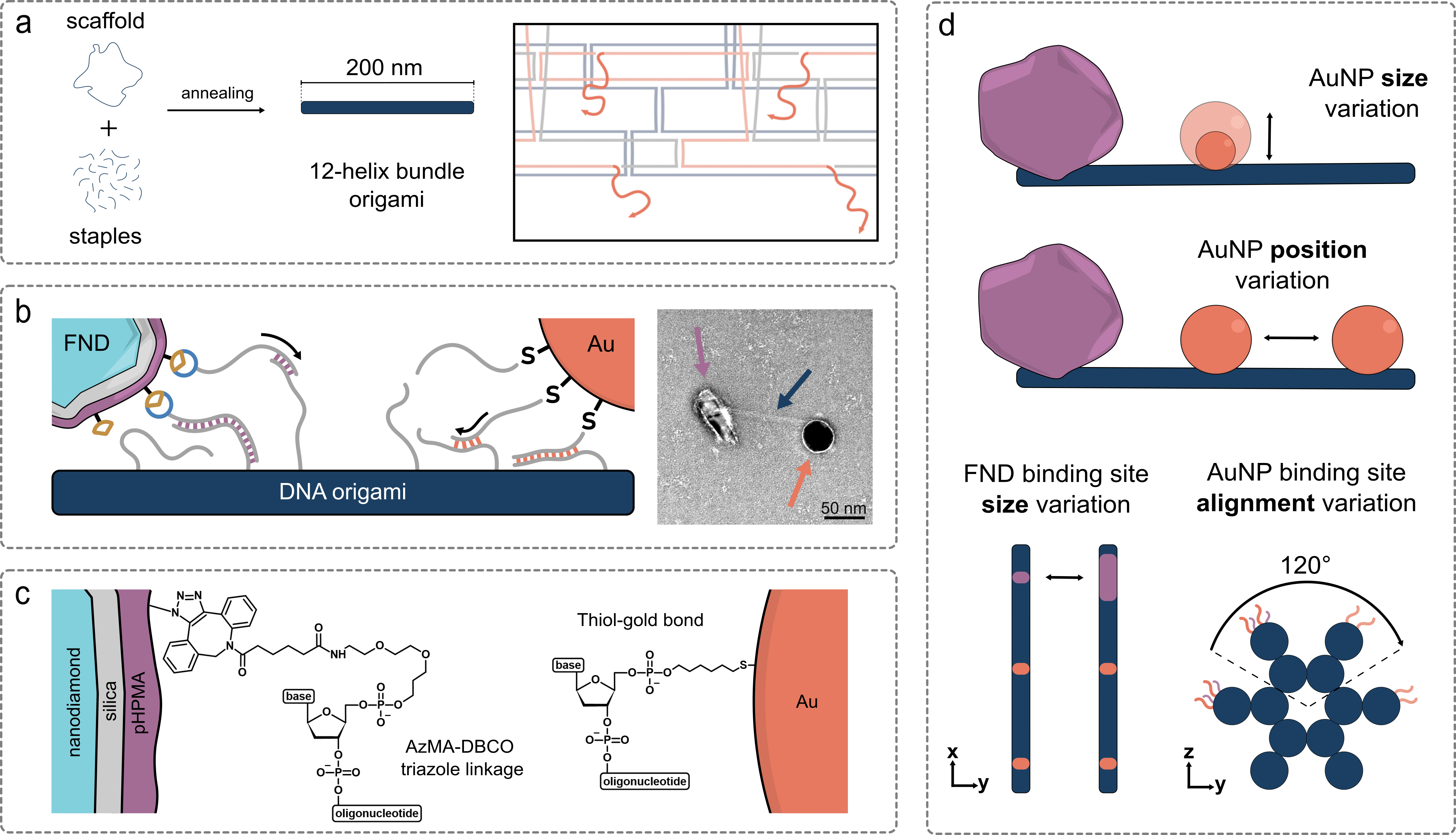}
    \caption{Overview of process steps towards heterogeneous nanoparticle assemblies. (a) Synthesis of DNA origami 12-helix bundles \textit{via} single stranded DNA scaffold and complementary DNA staples, resulting in DNA nanostructures containing addressable modifications in the form of ssDNA overhangs (inset). (b) Illustration of the binding mechanism of different nanoparticles onto the DNA origami. The TEM image shows a successfully assembled structure depicting the diamond nanoparticle (purple arrow), gold nanoparticle (orange arrow), and DNA origami (blue arrow). (c) Means of oligonucleotide surface modification of fluorescent nanodiamond and gold nanoparticle. (d) Overview of configurations with altering nanoparticle size (top), distance between the diamond and the gold nanoparticle (middle), and schematics of strategies to ensure robust particle assembly (bottom) by extending the binding site and by varying the respective position of the binding sites on the origami to obtain assemblies aligned along the long axis of the 12-helix bundle or shifted by 120\,\textdegree\, in respect to each other.}
    \label{fig:fig1}
\end{figure}

Decoration of DNA origami nanostructures with gold and diamond nanoparticles is based on base pairing of single-stranded overhangs on the origami structure with complementary oligonucleotides on the nanoparticle (Figure\,\ref{fig:fig1}a, b). 
A DNA origami-based platform is designed to position the nanoparticles. The structure is based on a well-established rod-shaped DNA nanostructure, the 12-helix bundle (12HB), which has been adapted to accommodate heterogeneous nanoparticles at defined distances. With an approximate length of 200\,nm and a diameter of $\sim$10\,nm, the high-aspect-ratio structure allows the controlled placement of multiple nanoparticles \cite{roller_hotspot-mediated_2017,gur_toward_2016}.
Three nanoparticle binding sites are typically incorporated, each consisting of four single-stranded DNA (ssDNA) overhangs with specific sequence, as illustrated in Fig.\,\ref{fig:fig1}a. This approach ensures that only nanoparticles functionalized with complementary ssDNA sequences can bind at predetermined locations (Figure \ref{fig:fig1}b). The individual binding sites are spaced approximately 70\,nm apart along the length of the DNA origami.

To facilitate nanoparticle attachment to the DNA nanostructure, each type of nanoparticle is functionalized with complementary oligonucleotides. AuNPs are conjugated with monothiol-modified oligonucleotides, as depicted in Figure\,\ref{fig:fig1}c. \\
In contrast to AuNPs, which can be directly functionalized with thiolated DNA, NDs require a more sophisticated surface engineering strategy to achieve stable and spatially controlled DNA attachment. To create a densely packed array of DNA, NDs designed to be colloidally stable in electrolytes with high ionic strength are beneficial. As known for the attachment of DNA to other nanoparticles, high concentration of electrolytes can compensate for the Coulombic repulsion of the individual DNA chains, allowing their dense organization. We designed polymer-coated NDs functionalized with azides and decorated them with DNA using strain-promoted azide-alkyne cycloaddition (SPAAC) (Figure\,\ref{fig:fig1}c), as described below. 

\subsection{Preparation and properties of DNA-modified FNDs }
To install DNA oligonucleotides in an orientationally controlled manner on the surface, chemically selective approaches are required. The available strategies addressing the end-specific attachment of DNA oligonucleotides to FNDs involve non-covalent attachment using biotin-avidin interaction \cite{zhang_dna-based_2015} or covalent attachment using copper-free click reaction \cite{akiel_investigating_2016}.

Here we present an alternative, a robust pathway for the creation of DNA arrays on FND particles. FNDs are coated with an ultrathin silica layer, serving as a base for attachment of a biocompatible translucent polymer-coating containing azide groups. This surface termination ensures the excellent colloidal stability of the particles in high-ionic-strength electrolyte (4\,{\texttimes}\,PBS in this case) \cite{neburkova_coating_2017} required for charge compensation of the DNA chains and their dense lateral loading. The bioorthogonally reactive azide groups in the copolymer serve as covalent anchoring points, allowing for the bio-orthogonal attachment of DBCO-modified DNA strands \textit{via} SPAAC. The nature of the SPAAC reaction enables running it for a long time which leads to high yields of conjugation while ensuring defined orientation of DNA attachment.

Technically, the surface modification process is three-step and involves i) formation of an ultrathin silica coating, ii) growth of a copolymer layer consisting of \textit{N}-(2-hydroxypropyl)methacrylamide and \textit{N}-(3-azidopropyl)methacrylamide using the "grafting from" approach \cite{slegerova_designing_2015}, and iii) bio-orthogonal attachment of DBCO-modified DNA strands \textit{via} SPAAC (Fig.\,1c). The individual synthetic steps i-iii) were confirmed using XPS (Figure\,S1). The formation of the copolymer layer supported by the ultrathin silica coating is documented by the presence of additional N1s and Si2p peaks compared to the starting bare FNDs. The attached DNA manifested in appearance of the P2s peak. The formed surface architecture was further analyzed thermogravimetrically (Figure\,S2). The addition of the organic structures is reflected in gradual decrease of the FND content in the particle and appearance of the weight losses corresponding to the organic groups from silica coating, HPMA copolymer, and finally DNA. The temperatures of the individual decomposition processes remained basically unchanged, further confirming the gradual formation of the target architecture. 

The polymer coating created this way provided negatively charged, sterically stabilized colloidal nanoparticles showing hydrodynamic diameters of around 100\,nm in water after DNA attachment (Supplementary Table 1). The loading of DNA was estimated using a fluorescent assay. The load of 198\,$\pm$\,6 oligonucleotides per FND corresponds to a densely packed DNA array (for a detailed description of the assay and the calculation, see Supplementary Note 1). Nanoparticle tracking analysis of the FND-polymer-DNA conjugates (Figure\,S3) revealed a fairly monodisperse population of the particles and enabled estimation of their number-weighted concentration (which is proportional to the molar concentration). Different batches of the particles can thus be used for the origami assembly at the same molarity, which is important for reproducibility of the assembly conditions. 

\subsection{Assembly \& Versatility}
To ensure the robustness and reliability of the assembly process, we employ several attachment site design strategies to mitigate potential challenges arising from possible steric and electrostatic interactions between nanoparticles and the restricted accessibility of DNA due to the complex chemical functionalization of nanodiamonds (Figure\,\ref{fig:fig1}d). For assemblies containing only FNDs and DNA origami, we evaluate three configurations: a simple attachment site with four ssDNA overhangs ("regular"), an extended attachment site with 20 overhangs near the original binding site ("extended"), and two "regular" attachment sites positioned at opposite ends of the origami structure ("double AS"). For the assemblies containing gold and diamond nanoparticles, we investigated steric interactions by varying the positioning of the AuNP binding site relative to the long axis of the rod-like origami. This resulted in two distinct configurations: an "aligned" configuration, where AuNPs and FNDs are bound on the same helices, and a "shifted" configuration, where the AuNP binding site is rotated by 120\,\textdegree\, with respect to the FND binding site. 

In this section, we present a systematic optimization of the DNA-mediated assembly process. We use the term 'imaging yield' to quantify the number of successfully assembled nanoparticles after purification, calculated from AFM images of structures deposited on mica. Imaging yield represents the percentage of correctly assembled structures (example shown in Figure\,\ref{fig:fig2}d) within the sample. The assembly process produces a range of structures, including unintended assemblies. Figure\,\ref{fig:fig2}g highlights some of these cases, such as multiple origami binding to a single nanoparticle or nanoparticle clustering. We first focus on optimizing the binding of FNDs to the DNA origami before extending our analysis to the heterogeneous FND-AuNP assembly.

\subsubsection{Nanodiamond binding to DNA origami}
\begin{figure}[ht]
    \centering
    \includegraphics[width=\textwidth]{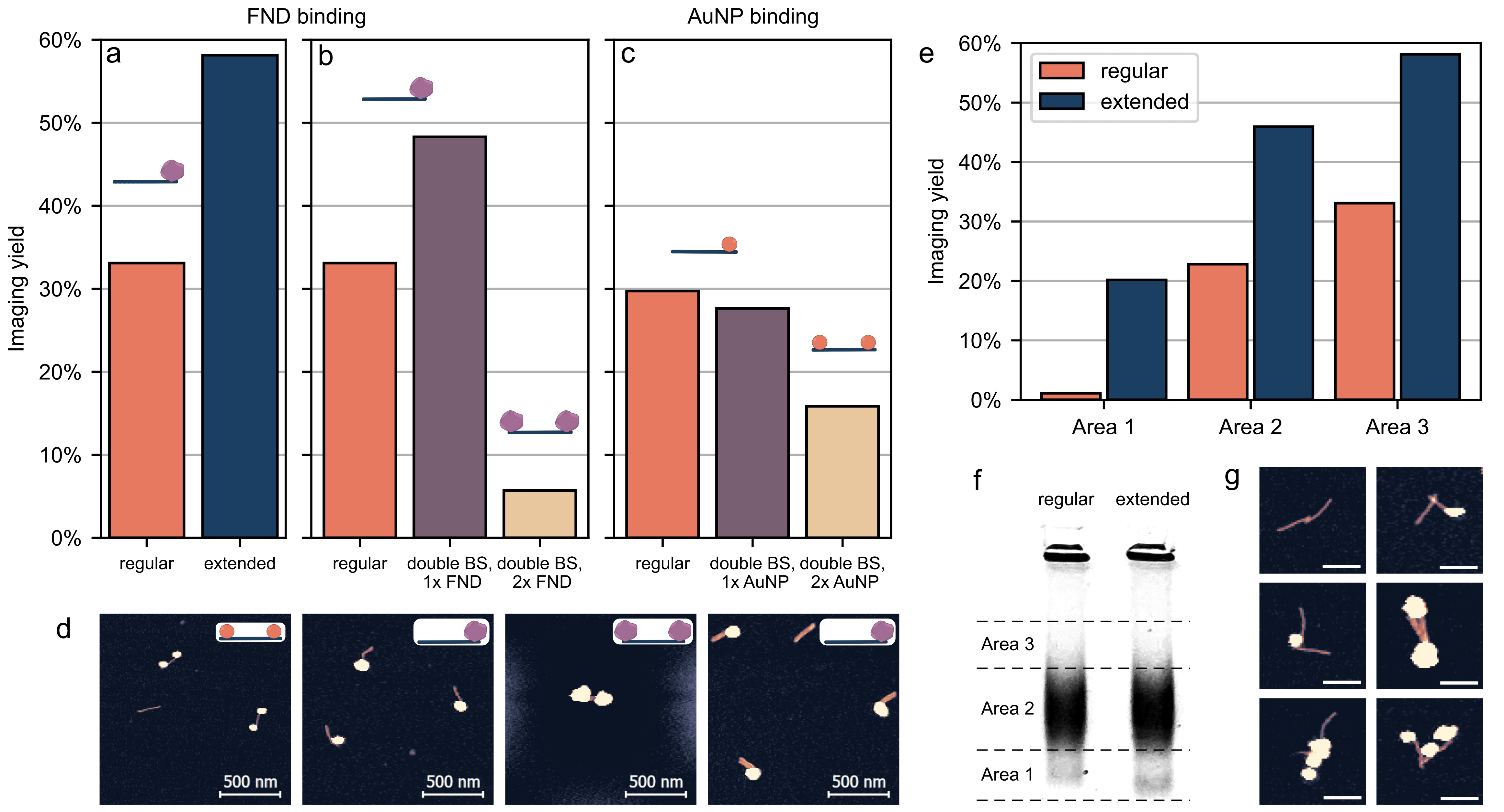}
    \caption{Binding yield of nanoparticles on the origami using different strategies. a) AFM imaging yields of FND bound to origami using either four staple elongations ("regular", orange) or 20 staple elongations ("extended", dark blue). b) Comparing FNDs bound to origami with one "regular" binding site and with two "regular" binding sites ("double BS"), resulting in one ("1\texttimes\,FND", dark purple) or two nanodiamonds ("2\texttimes\,FND", beige) being bound to the origami. c) Control experiment with binding of 40\,nm AuNP functionalized with the same sequence as the FNDs ((TTG)$_7$). (d) Example images of different structures recorded in AFM, the height scale is 5\,nm. (e) AFM imaging yield of samples containing FNDs bound to 12HB with "regular" and "extended" binding sites from different areas of the gel lane used for purification. Indicating that Area 3 - the top of the gel - contains the highest yield of desired structures. (f) Representative image of gel electrophoresis lanes containing DNA origami nanostructures and DNA-functionalized FNDs after undergoing incubation. Areas 1\,-\,3 depict the extracted portions of the lane that are further analyzed. The dark color represents the fluorescent signal from the nanodiamond. (g) Example images of failed assemblies, consisting of undesired binding of different components and clustering through particle aggregation of origami cross-linking, height scale is 5\,nm, scale bar indicates 200\,nm. }
    \label{fig:fig2}
\end{figure}

To bind the FNDs on the origami, fully assembled 12-helix bundles (12HB) are mixed with FNDs at a 1:2 molar ratio and heated slightly above the melting temperature of the linking sequence, followed by slow annealing to room temperature to ensure proper binding. Figure\,\ref{fig:fig2}a-c compares the binding efficacy of different configurations, including AuNPs as a reference (Fig.\,\ref{fig:fig2}c). The 40\,nm AuNPs undergo the same assembly protocol as the FNDs.
Three binding configurations are evaluated: "regular" (four complementary ssDNA overhangs), "extended" (20 overhangs), and "double AS" (two binding sites per origami, each containing four complementary ssDNA overhangs).

AFM image analysis (Figure\,\ref{fig:fig2}a\,-\,c) reveals differences in binding efficiency across these configurations. The binding efficiency of AuNPs and FNDs in the "regular" configuration is comparable ($\sim$30\,\%), suggesting that DNA coverage on the nanoparticle surface does not significantly affect assembly efficiency. This conclusion is drawn from the fact that AuNPs  approximately have twice the oligonucleotide surface density of FNDs when comparing the theoretical maximum loading of oligonucleotides on 40\,nm AuNPs (n\,=\,430\,$\pm$\,10) \cite{hill_role_2009} and the measured DNA surface functionalization on FNDs (n = 198\,$\pm$\,6).
Increasing the number of overhangs from four ("regular" configuration) to 20 ("extended" configuration) significantly improves imaging yield from 33\,\% to 58\,\% (Figure\,\ref{fig:fig2}a). The elongation of the binding site may come with reduced binding precision, as the attachment site extends from 6\,nm to $\sim$30\,nm along the origami axis. This can lead to positional offsets from the desired positioning relative to the gold nanoparticles in subsequent steps. Despite this, AFM images (Figure\,\ref{fig:fig2}d and Supplementary Figure\,S5) indicate that major deviations are rare, with structures mostly maintaining expected configurations when compared to FND-origami structures with the "regular" configuration.
Doubling the number of binding sites results in distinct behaviors (Fig.\,\ref{fig:fig2}b,\,c). We evaluated two resulting assemblies: both binding sites occupied by the nanoparticles ("double BS, 2\texttimes\,FND/AuNP"), and a single binding site occupied by the nanoparticle ("double BS, 1\texttimes\,FND/AuNP"). In the case of AuNPs, two-particle structures account for $\sim$15\,\% of the assemblies, while single-particle assemblies remain at around 30\,\%. In contrast, FND single-particle binding increases by nearly 50\,\% relative to the "regular" configuration, but two-particle assemblies remain low ($\sim$5\,\%). This discrepancy may arise from differences in particle size, shape, and diffusivity, affecting binding dynamics. In combination with the overall lower oligonucleotide density of the particle surface, a difference in binding behavior can be expected. 

Values in Fig.\,\ref{fig:fig2}a\,-\,c represent gel-optimized imaging yields. After the hybridization process, the assembled structures are separated from excess origami and nanoparticles \textit{via} agarose gel electrophoresis. The imaging yield is determined as the maximum observed fraction across three primary regions of the gel, with the highest yield consistently found in the top area of the gel (Area 3) for FND-bound structures (Fig.\,\ref{fig:fig2}e). The gel exhibits three primary regions: Area 1, which contains mostly excess origami, Area 2, which shows strong FND fluorescence with a broad distribution due to the relative polydispersity of the FNDs, and Area 3, which contains only a tail of the FND band but does not exhibit any distinct fluorescence signal of DNA (Fig.\,\ref{fig:fig2}f). The slightly shifted bands between the "regular" and "extended" configurations indicate differences in overhang numbers. A comprehensive dataset across all configurations and gel sections is provided in Supplementary Note 3.

\subsubsection{Heterogeneous particle assembly}
\begin{figure}[ht]
    \centering
    \includegraphics[width=\textwidth]{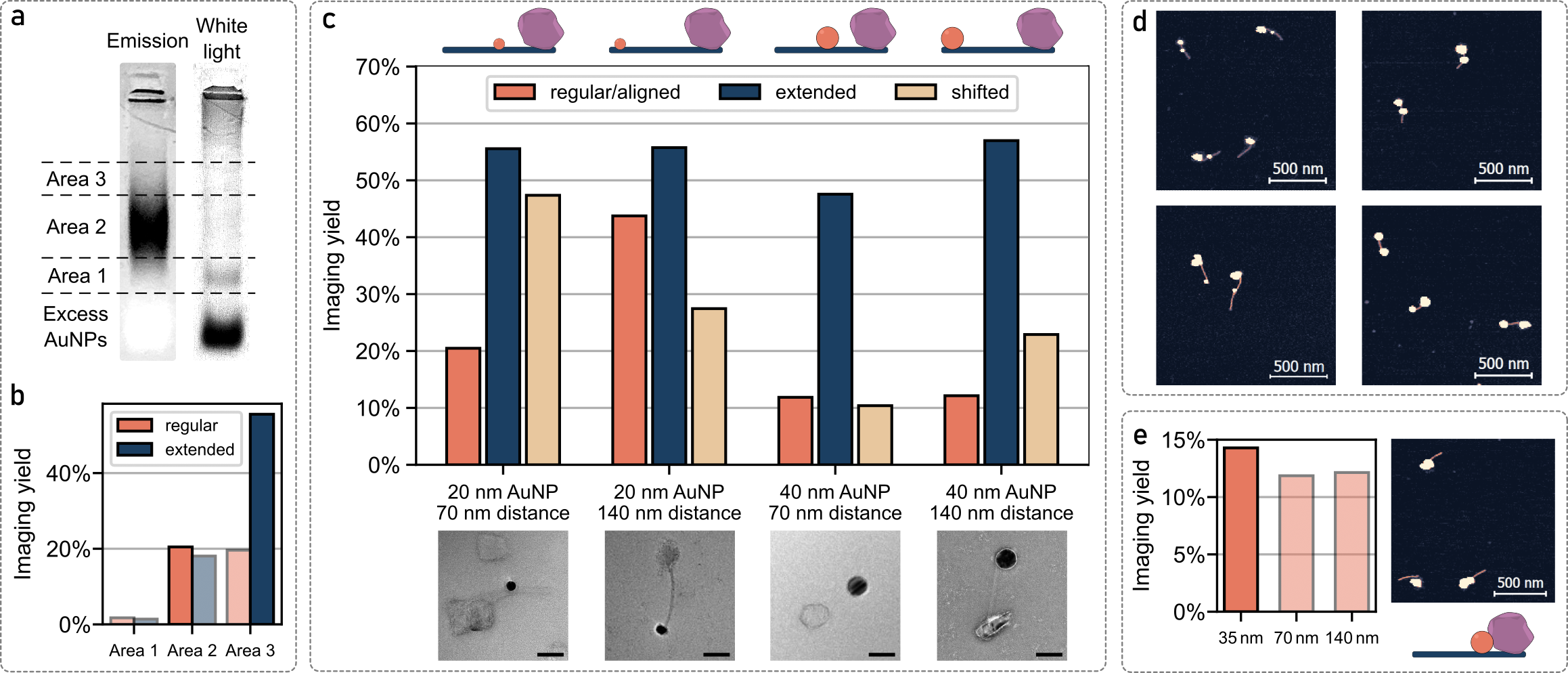}
    \caption{Forming assemblies of gold and diamond nanoparticles. (a) Image of the agarose gel of a typical AuNP-FND-origami sample after purification under two imaging conditions. Under blue light, the strong FND emission is visible as a thick band in the center of the lane, whereas under white light illumination, the AuNP bands are visible. Lines indicate separation in three distinct areas for AFM analysis. (b) Results of AFM analysis according to cut bands for a sample containing 40\,nm AuNP and FND on origami containing "regular" and "extended" binding site configurations for FND. Highlighted bars represent gel-optimized yields used in further comparison. (c) Overview of imaging yields for tested configurations containing 20\,nm and 40\,nm AuNP with different interparticle distances. Illustrations above the barplot represent respective structures with components to scale. The assembly with "extended" binding site has dominantly the highest yield across all configurations. Additional TEM images confirm the successful assembly of desired structures, scale bar represents 50\,nm. (d) Representative AFM images as used for analysis, showing 20\,nm AuNP assemblies with 70\,nm and 140\,nm distance between particles in the left column (top to bottom) respectively, and 40\,nm AuNP with the same particle-particle distance configurations in the right column. (e) Comparison of imaging yield after further reduction in interparticle distance down to 35\,nm. Gel-optimized imaging yields of configurations containing one 40\,nm AuNP with varying binding site distance to the FND as indicated on the x-axis. AFM snapshot showing representative structures used in analysis.}
    \label{fig:fig3}
\end{figure}

After successfully binding FNDs to the 12HB structure, the next step involves assembling FNDs and AuNPs together to investigate potential modulation of FND emission behavior. To achieve this, FNDs are first attached to the origami structure as described, followed by the addition of freshly functionalized AuNPs in a 1:1 molar ratio to the unpurified mixture. After incubation at room temperature, the samples are purified \textit{via} agarose gel electrophoresis. The presence of AuNPs alters the gel lane pattern: the stained gel reveals FND emission but no longer shows an excess DNA origami band in Area 1, as illustrated in Figure\,\ref{fig:fig3}a. Instead, the excess origami band is replaced by a lower-positioned excess AuNP band, along with a second band, visible to the naked eye, corresponding to origami-AuNP structures.

Moving further up the gel, most of the desired assemblies are found in Area 2 and Area 3, as illustrated in Fig.\,\ref{fig:fig3}b. Comparing the imaging yield across these regions for the "regular" and "extended" configurations reveals differences in electrophoretic mobility based on nanoparticle attachment style. In the "regular" configuration, structures are evenly distributed between Area 2 and Area 3, whereas the "extended" configuration exhibits a comparable yield in Area 2 but a sharp increase in Area 3 (Fig.\,3b). 

A broader analysis of gel-optimized imaging yields across different configurations and nanoparticle combinations reveals distinct trends (Fig.\,\ref{fig:fig3}c) showing the resulting yields and example TEM images of the assemblies. Characteristic AFM images of the four different binding site positions and AuNP size configurations are found in Figure\,\ref{fig:fig3}d. 

In the "regular" configuration (orange), 20\,nm AuNPs exhibit a significant increase in imaging yield with increasing distance from the FND. However, when the AuNP size is increased to 40\,nm, the overall number of assembled structures decreases, with minimal variation across different attachment site distances. Several factors contribute to this trend. First, the linking sequence between origami and AuNP is (TTT)$_7$, which differs from that used in the origami-AuNP assemblies shown in Fig.\,\ref{fig:fig2}a. The lower melting temperature of this sequence enhances AuNP binding efficiency (data not shown), making FNDs less likely to bind. Secondly, considering 20\,nm AuNPs, reduced steric hindrance allows efficient assembly, with longer interparticle distances further improving yield. However, for 40\,nm AuNPs, steric interactions become more pronounced, lowering the overall assembly yield even at longer distances. 

To address both of these factors and assess their importance on the assembly yield, we test two strategies: i) Increasing the number of elongations for nanodiamond from four ("regular") to 20 ("extended") and ii) shifting the nanoparticle binding position on the origami to reduce the steric interactions. 

Comparing the "regular" and "extended" configurations reveals a stark contrast in assembly yield. Across all AuNP sizes and interparticle distances, the "extended" configuration consistently achieves yields above 47\,\%, supporting the hypothesis that FND binding is a limiting factor in the assembly process. The increased number of binding sites compensates for competing effects, leading to a nanoparticle size- and distance-independent assembly. 

Probing the impact of nanoparticle orientation along the long axis of the 12HB yields mixed results indicating that the steric interaction is not the dominant driving factor for the successful assembly. In the "shifted" configuration, where the AuNP binding site is rotated, the close configuration with 20\,nm AuNPs shows a strong yield increase compared to the "regular" arrangement. However, at greater interparticle distances, the yield drops below that of the "regular" configuration. This may indicate the twisting of the 12-HB origami, as reported by cryoEM studies \cite{kube_revealing_2020}, making the configuration where nanoparticles are arranged in a straight line in the origami design appear "shifted" in real assembly and vice versa. For 40\,nm AuNPs, no pronounced differences are observed between the "aligned" and "shifted" conformations. The increased yield at the 140\,nm interparticle distance configuration of the "shifted" sample does not match the profound improvements observed for the extended sites. 

To investigate the modulation of the photoluminescence of NV centers by plasmonic nanoparticles, we added one more sample configuration to the set. An extreme case with reduced distance of the nanoparticle binding sites to 35\,nm (Fig.\,\ref{fig:fig3}e), making the nanoparticles almost touching one another. This experiment was conducted using only the "regular" configuration with 40\,nm AuNPs, which resulted in a yield comparable to that of other distance configurations. AFM image analysis becomes more challenging in this case due to tip convolution effects.

\subsection{Plasmonic coupling of NV centers}
\begin{figure}[ht]
    \centering
    \includegraphics[width=1\linewidth]{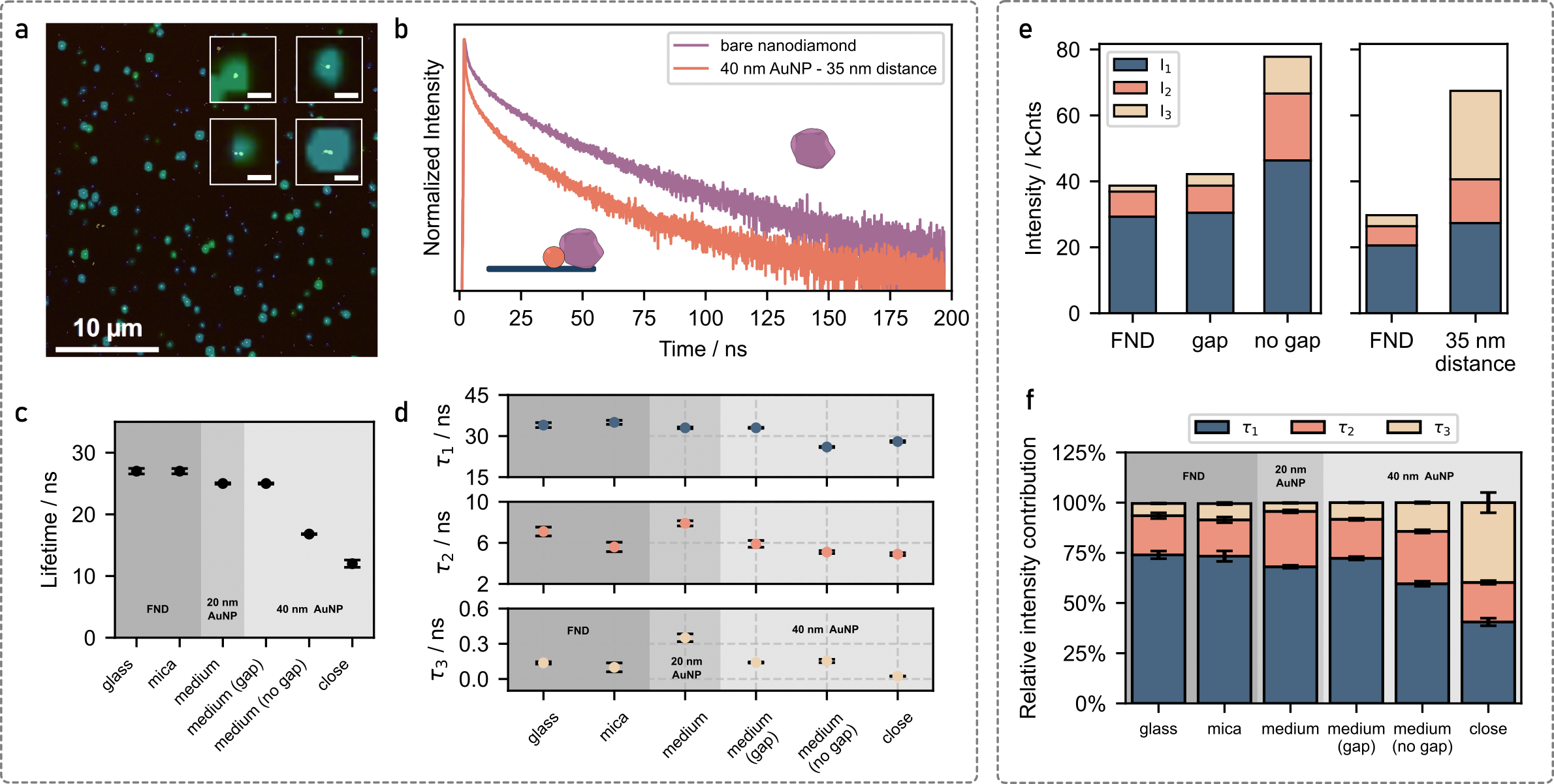}
    \caption{Modulation of the photoluminescence of NV centers changes based on the distance between the FND and AuNP. (a) FLIM/AFM overlay image consisting of intensity and lifetime data paired with the same area imaged \textit{via} AFM. Insets show example structures with their respective FLIM signal, scale bar represents 500\,nm. (b) Example decays of non-functionalized nanodiamonds on the composite substrate and assemblies of short distance (35\,nm center-to-center) AuNP-FND configuration, showing the overall lifetime reduction. (c) Intensity-weighted average lifetime of bare FNDs on glass and on composite mica substrate (dark grey area), 20\,nm AuNP-FND assemblies (70\,nm center-to-center distance, medium grey area), and 40\,nm AuNP-FND assemblies with decreasing distance between the two particles, resulting in decreased average lifetime (light grey area). "Medium" describes the 70\,nm interparticle distance configuration and "close" the 35\,nm configuration. (d) Lifetime components of each particle and assembly configuration showing decrease in the lifetime of the longest components in "medium (no gap)" and "close" configuration samples. The shortest lifetime components are on the resolution limit of the FLIM and their exact values are non-conclusive. (e) Particle number-normalized absolute intensity of the PL emission for samples of different 40\,nm AuNP configurations. The FND reference intensity is extracted from the signals of single FNDs (with no AuNP) present in the sample from the same measurement by correlating with the AFM image. It varies from sample to sample due to the mica sheets in composite substrate (f) Relative intensity contribution of individual lifetime components for all particle and assembly configurations showing the increasing contribution of the fastest component with reducing distance.}
    \label{fig:fig4}
\end{figure}

Building on our optimized assembly strategy, we systematically explore how the spatial arrangement of plasmonic nanoparticles influences the photoluminescence of NV centers, providing insight into the mechanisms of plasmon-NV coupling. Collective excitations of electrons in plasmonic nanoparticles lead to a dramatic increase in the local density of optical states near the nanoparticle surface \cite{novotny_principles_2012, le_ru_principles_2009, anger_enhancement_2006, butet_optical_2015}. When a fluorescent molecule (in our case NV center) is placed in proximity to a plasmonic nanoparticle, all quantum processes contributing to fluorescence - including absorption, dephasing, excited-state transitions, and emission - may be altered. The extent and nature of these modifications depend on several factors: i) the intrinsic photophysical properties of the fluorophore, ii) the resonance energies of the plasmonic oscillations, and iii) the spatial arrangement of the fluorophore and nanoparticle, including their separation distance, relative orientation, and the optical properties of the surrounding medium.

Fluorescence lifetime imaging (FLIM) can probe plasmonic interactions, as it provides quantitative information on the modification of the excited-state dynamics by the local environment. This technique offers two key characteristics: the total photon emission counts and the temporal distribution of excited-state lifetimes before emitting a photon, providing insight on how the coupling influences radiative and non-radiative decay pathways, energy transfer processes, and emission efficiency. 

We used FLIM to analyze changes in emission of the NV centers due to the presence of plasmonic nanoparticles on the level of single structures. To distinguish desired assemblies from single FNDs, clusters, and other undesired structures, we correlate the FLIM measurement with highly resolved AFM images. To ensure sufficient quality of the AFM image and enable registration with FLIM images, we deposit the samples on a flake of mica fixed on a lithographically labeled coverslip. This way, each individual structure measured in the AFM can be assigned a distinct signal in FLIM as illustrated in Figure\,\ref{fig:fig4}a. We threshold the signal of the desired type of structure and analyze the decay signal. An example of the decay of nanodiamonds and the closest AuNP-FND assembly is shown in Figure\,\ref{fig:fig4}b with distinctly different fluorescence decay behavior. 

We analyzed six distinct types of structures: 1) nanodiamonds on glass substrate, 2) nanodiamonds on a mica-glass substrate, 3) FND-AuNP assemblies with 20\,nm gold nanoparticles, with a center-to-center distance of 70\,nm (labeled "20\,nm AuNP medium"), 4) FND-AuNP assemblies with 40\,nm gold nanoparticles, with a center-to-center distance of 70\,nm, where a clear gap between the nanoparticles is visible in the AFM image (labeled "40\,nm AuNP medium (gap)"). In this case, the spacing between the nanoparticles is estimated to be 30\,-\,40\,nm according to TEM. 5) The same assembly design as in case 4, but with no apparent gap between the nanoparticles (see insets in Figure\,\ref{fig:fig4}a), resulting in a spacing of less than 20\,nm (labeled "40\,nm AuNP medium (no gap)"). The variation in spacing within a single assembly design is due to the heterogeneity of the nanodiamond sizes and, to a limited extent, to the use of the "extended" binding site for the nanodiamonds, which is approximately 30\,nm in length. 6) FND-AuNP assemblies with 40\,nm gold nanoparticles and a center-to-center distance of 35\,nm, labeled "40\,nm AuNP close." The emission intensity comparison was done by evaluating single unbound nanodiamonds and AuNP-FND assemblies from the same sample, to avoid inaccuracies caused by differences in substrate heterogeneity. 

The measured fluorescence decays of selected structures are deconvoluted into their individual lifetime components by fitting with a three-exponential model (Fig.\,4d), Through this analysis the intensity-weighted average lifetime is extracted.
Figure\,\ref{fig:fig4}c depicts these intensity-weighted average lifetime of analyzed structures. First, influences of the substrate are probed by measuring bare nanodiamonds on glass coverslips and on the composite substrate. No difference can be detected with both substrates resulting in an average lifetime of 27.0\,ns\,$\pm$ 0.5\,ns (n = 12) on glass and 27.0\,ns\,$\pm$ 0.4\,ns (n = 13) on the composite substrate. These lifetimes are close to reported literature values of around 25\,ns \cite{beveratos_nonclassical_2001, inam_emission_2013}. These differences can be accounted for with general heterogeneity of the FNDs, relative position of the NV center in regards to the interface and random dipole orientation. Nitrogen vacancy center-containing nanodiamonds are also known to be sensitive in regards to their emissive properties depending on size and the irradiation parameters used in their fabrication. \cite{chung_particle_2007, laube_controlling_2019}. 20\,nm and 40\,nm AuNPs affect the fluorescence lifetime of the FNDs marginally in the 70\,nm distance configuration. Both exhibit an average lifetime of 25\,ns (n$_{20\,nm}$ = 12, n$_{40\,nm}$ = 10) which is within the normal range for the average FND lifetime reported prior. 

Examining structures containing only 40\,nm gold nanoparticles reveals a clear dependence on inter-particle distance. Assemblies in which a visible gap is present between the nanoparticles exhibit an average fluorescence lifetime comparable to that of bare nanodiamonds ($\tau_{avg}$ = 25.0 $\pm$ 0.1\,ns, n = 10). In contrast, structures with no apparent gap show a significant reduction in lifetime, decreasing to $\tau_{avg}$ = 16.8 $\pm$ 0.1\,ns (n = 14). This trend continues in the closest-spaced assemblies, where the fluorescence lifetime is further reduced to $\tau_{avg}$ = 12.0 $\pm$ 0.6\,ns (n = 12), indicating a progressive modulation of NV center emission properties with decreasing separation between the gold nanoparticle and nanodiamond. 

The individual components of the fluorescence decay are visualized in Figure\,\ref{fig:fig4}d. $\tau_1$ represents the longest component with around 34\,ns for bare FNDs and assemblies containing 20 and 40\,nm AuNPs with a medium interparticle distance. Bringing AuNP and FND closer together results in a reduction of this component to around 27\,ns. The second component $\tau_2$ shows fluctuation around 6\,ns. The shortest component $\tau_3$ is around 300\,ps which is on the border of the resolution limit of the FLIM setup. While this component accounts for a minor contribution in bare nanodiamonds and medium distance AuNP-FND assemblies, it becomes more pronounced in the assemblies with closer distances (no gap) and close assembly (Figure\,\ref{fig:fig4}f). 

To quantify the modulation of NV center emission by plasmonic nanoparticles, we calculated the Purcell factor ($F_P$), which describes the enhancement of the total decay rate altered by the modified local density of optical states of the environment. Using the relation $F_P = \tau_{\text{bare}} / \tau_{\text{sample}}$, we get $F_P = 1.61$ for the middle distance (no gap) assemblies and $F_P = 2.25$ for the close configuration. Both samples exhibit an increase in the fluorescence emission (Figure\,\ref{fig:fig4}e). However, the fluorescence intensity increase does not scale proportionally with $F_P$, suggesting that while radiative enhancement contributes significantly, non-radiative channels also play a role. 

The observed reduction in average lifetime can be interpreted within the framework of theoretical models describing quantum processes in plasmon-coupled NV centers \cite{hapuarachchi_nv-plasmonics_2022}. These models predict that decay rates can be enhanced by up to a factor of 7.5 when the NV center is optimally aligned with the plasmonic modes, suggesting that even greater enhancements are achievable under ideal conditions. In our case the NV orientation was random, meaning the coupling efficiency was not maximized. 

To disentangle the radiative and non-radiative contributions, we estimated the relative quantum efficiency ($\eta_{\text{rel}}$) from the normalized fluorescence intensity and lifetime data. We found that at the middle distance, the radiative decay rate was enhanced close to the total Purcell factor ($\eta_{\text{rel-middle}} = 1.22$), indicating predominantly radiative effects. At close configuration, the fluorescence intensity increase was less pronounced despite a stronger lifetime reduction yielding $\eta_{\text{rel-close}} = 1.04$, significantly lower than the calculated Purcell factor. This implies a competing effect from non-radiative processes. 

Interesting is the increasing contribution of the shortest lifetime component in the FLIM analysis of close-distance assemblies. While this component accounts for only $\sim5\,-\,10\,\%$ in bare nanodiamonds (which can be attributed to scattering of the diamond crystal itself), it rises to $\sim 18\,\%$ in the middle configuration and nearly $40\,\%$ in the short-distance assemblies. It suggests that the enhancement mechanism is not uniformly distributed across all decay pathways and an ultrafast decay channel emerges. It points toward a selective enhancement of a specific decay component as has been reported in systems where strong coupling to localized plasmonic modes occurs \cite{mueller_deep_2020}. In these cases, the local density of optical states is modified to preferentially enhance emission through certain transitions, leading to an increase of the fastest decay component. Energy redistribution between radiative and non-radiative channels can also cause this effect. If the plasmonic field enhances emission from a particular dipole transition in the NV center more effectively than others, an increase in the relative contribution of the fastest decay component can occur. 

The enhancement of the shortest decay pathway is of particular importance for technologies that require fast light-matter interactions. For example, it can be beneficial for accelerating spin-state readout rates for quantum sensing \cite{he_direct_2024,irber_robust_2021}. In integrated photonics, faster decay pathways are desirable to decrease jitter in single photon sources, where reduced excited state lifetimes minimize timing uncertainty \cite{benedikter_cavity-enhanced_2017, korzh_demonstration_2020}. Further modification of the geometry of the assembly (for example the gap between the nanoparticles) might tune the balance between the radiative and non-radiative decay rates for desired application direction. While applications in quantum optics seek to maximize the radiative decays, non-radiative decays can be beneficial for energy conversion. 

\section{Conclusion}
In this work, we expand the capabilities of DNA origami-based nanophotonics to include NV centers in diamond, demonstrating a scalable and precise method for assembling nanodiamond-gold hybrid structures. Our strategy overcomes previous limitations in functionalizing NDs for DNA-directed assembly, enabling high-yield and spatially controlled positioning. By systematically probing fluorescence lifetime modifications as a function of interparticle distance, we demonstrate the interplay of plasmon-mediated effects on the photoluminescence of NV centers. We show that at short interparticle distances, the fastest decay pathway becomes the dominant contribution to the photoluminescence enhancement, while at medium distances the decay pathways are enhanced uniformly. The change in emission pathways indicates contribution of both radiative and non-radiative processes. This highlights the ability to tune the decay pathways of NV center emission by controlling the geometry of the assembly. The high structural yield and reproducibility of our approach underscore its potential for engineering complex nanoscale optical systems. In our future work we will use the developed devices to study how the mechanism of plasmonic enhancement affects the position of their projection to construct sensors of molecular dynamics.

\section{Methods\label{sec:Methods}}
\subsection{Chemicals}
M13mp18 scaffold was purchased from tilibit Nanosystems (Germany). The oligonucleotide for FND functionalization and staple strands for 12-helix bundles were purchased from Metabion (Metabion International AG, Germany) with HPLC purification; the oligonucleotide for FND surface loading determination (hereafter referred to as cDNA-ATTO488; see Supplementary Note 1 for the sequence) was purchased from IDT (Integrated DNA Technologies, USA) with HPLC purification. Nanodiamonds (MSY 0\,-\,0.05) were obtained from Microdiamant (Switzerland). TAE buffer (40\,\texttimes) was purchased from Promega (USA). Magnesium chloride (MgCl$_2$), sodium chloride (NaCl), tris(hydroxymethyl)aminomethane (TRIS; pH\,=\,8.0), bis(p-sulfonatophenyl)phenylphosphine (BSPP), tris(2-carboxyethyl)phosphine (TCEP), sodium dodecyl sulphate (SDS), glycerol, polyvinylpyrrolidone (PVP, M$_w$\,=\,10,000\,Da), tetraethylorthosilicate (TEOS), (trimethoxysilyl)propyl methacrylate (TMSPMC), azobisisobutyronitrile (AIBN), N-(2-hydroxypropyl)-methacrylamide (HPMA), (4-(2-hydroxyethyl)-1-piperazineethanesulfonic acid) (HEPES; pH\,=\,7.4), phosphate-buffered saline (PBS), and RNase-free water were purchased from Merck Life Science (Merck, Czech Republic). 
Sulfuric acid (H$_2$SO$_4$), nitric acid (HNO$_3$), perchloric acid (HClO$_4$), sodium hydroxide (NaOH), hydrochloric acid (HCl), and ammonia solution were purchased from Lach-ner (Czech Republic). 
HPLC solvents such as ethanol, methanol, ethyl acetate (EtOAc) and dimethyl sulfoxide (DMSO) were purchased from VWR International (VWR, Czech Republic). 
Milli-Q water was used to prepare all the solutions.

\subsection{DNA origami synthesis}
The 12-helix bundle was redesigned using caDNAno software \cite{douglas_rapid_2009}. The list of staple sequences and can be found in Supplementary Note 5. Origami nanostructures (12-helix bundles) are synthesized by mixing M13mp18 scaffold with the respective staple strands in a 1:10 molar ratio.  Modified staple strands for binding nanoparticles are added separately in a 1:15 ratio to ensure their incorporation. The mixture is brought to 1\,{\texttimes}\,TAE, 15\,mM MgCl$_2$ and incubated in Biometra TAdvanced thermocycler (Jena Analytik, Germany) for around 25\,h according to established protocols \cite{scheckenbach_selfregeneration_2021}. After completion, the structures are purified by five rounds of filtration in 100\,kDa Amicon\textregistered\, MWCO Ultra centrifugal filters (Merck Milipore, Germany). After recovery of the retentate, the concentration is determined spectrophotometrically with an NP80 NanoPhotometer (Implen, Germany).

\subsection{AuNP functionalization}
The procedure is based on a modified protocol by G\"{u}r et al. \cite{gur_toward_2016}. To increase the particle concentration, the AuNPs are first mixed with BSPP to reach a concentration of 2.5\,mM. The mixture is shaken overnight at room temperature. The solution is then centrifuged for 30 minutes at 15,000\,{\texttimes}\,g. The supernatant is discarded, and the nanoparticles are redissolved in 1\,mL freshly prepared 2.5\,mM BSPP and 1\,mL MeOH. The solution is then again centrifuged at 15,000\,{\texttimes}\,g. The supernatant is discarded and 1\,mL of 2.5\,mM BSPP solution is added. The absorbance of the solution at 450\,nm and the corresponding plasmon peak is measured on the same nanophotometer as the origami structures and the concentration is determined by calculating the mean value of the concentrations at the two wavelengths using differently sourced extinction coefficients \cite{haiss_determination_2007, dolinnyi_extinction_2017}. Monothiolated ssDNA strands with a T$_{21}$ sequence are incubated with TCEP for one hour in a 1:250 molar ratio to reduce disulfide bonds. The solution is then added without further purification to the concentrated AuNP solution in varying ratios according to the size of the NPs \cite{hill_role_2009}. Salt-aging is performed to ensure sufficient binding of oligonucleotides to the particle surface. To prevent aggregation of particles $\ge$\,40\,nm, SDS is added to a concentration of 0.2\,\%\,w/v. The mixture is brought to 0.75 M NaCl by adding 5\,M NaCl solution in ten steps. After each addition, the solution is extensively vortexed, sonicated for 10\,s, and incubated for 20\,min. Afterward, the solution is left on the shaker overnight at room temperature. The solution is then purified to remove excess unbound linkers by ultrafiltration in 100\,kDa Amicon\textregistered\, MWCO filters. Five rounds of centrifugation at 10,000\,{\texttimes}\,g for 5\,min are performed and after each round 400\,{\textmu}L of 1\,{\texttimes}\,TAE is added to the filter. After the last round, the retentate is recovered from the filter and the concentration is determined as described before. The functionalized nanoparticles are used immediately after purification.

\subsection{Fluorescent nanodiamond functionalization}
\subsubsection{Oxidation of nanodiamonds (NDs)}
1.5\,g of nanodiamonds was ground in an agate mortar with pestle and oxidized by air in a tube furnace for 6\,h at 540\,\textdegree C, mixing it every 20\,min. After cooling, NDs were transferred into a PTFE container and 6\,mL of H$_2$SO$_4$, HClO$_4$, and HNO$_3$ was added. The mixture was bath sonicated for 10\,min and then heated to 90\,\textdegree C for 48\,h with stirring. After reaction, the mixture was cooled to room temperature and divided into two Falcon tubes placed in ice. The Falcon tubes were centrifuged using a swinging rotor at 3,969\,\texttimes\,g for 15\,min (25\,\textdegree C). The supernatant was removed and the pellets were resuspended in water. This washing step was repeated once with 1\,M NaOH solution, once with water, and again with 1\,M HCl solution and with water again. Then, ND solution was transferred into six centrifuge tubes and centrifuged six times with water at 30,000\,{\texttimes}\,g for 25\,min (25\,\textdegree C). After the last centrifugation, the colloidal solution was lyophilized providing approximately 1.0\,g of grey ND powder. 

\subsubsection{Preparation of fluorescent nanodiamonds (FNDs)}
FNDs were prepared from the purified ND powder described above. 1.0\,g of ND powder was placed on a thick aluminium target. The sample was irradiated using an electron beam (80\,h, 6.6\,MeV, 1.25$\cdot 10^{19}$\,particles/cm$^2$). Post-irradiation, the sample was annealed at 900\,\textdegree C for 1\,h under argon atmosphere. Then the sample was processed in the same manner as before (air oxidation at 540\,\textdegree C, 5\,h; three-acid oxidation at 90\,\textdegree C, 48\,h) and lyophilized. Approximately 750\,mg of FNDs was obtained.

\subsubsection{Silication of FNDs}
6\,mg of FNDs were combined with 3\,mL of water in a plastic tube. The suspension was sonicated for 60\,min using cup horn sonication (pulse on: 1\,s, pulse off: 1\,s, amplitude 40\,\%). Separately, 16.8\,mg of PVP was dissolved in 15\,mL of water in a 50\,mL Erlenmeyer flask. This solution was bath sonicated for 15\,min. The FND colloidal solution was then gradually added to the PVP solution under stirring. The mixture was stirred at room temperature (RT) for 24\,h.
The solution was evenly divided between two Falcon tubes and centrifuged at 40,000\,{\texttimes}\,g for 60\,min (10\,\textdegree C). After centrifugation, the supernatant was removed, pellets were resuspended in a small amount of water and transferred into 1.5\,mL Eppendorf tubes for second centrifugation at 30,000\,{\texttimes}\,g for 30\,min (10\,\textdegree C). Following the removal of the supernatant, the remaining pellet in each tube was resuspended in a small amount of water (total volume 450\,{\textmu}L). The FND colloidal solution was added to 6\,mL of ethanol followed by addition of 45\,{\textmu}L of TEOS (0.20\,mmol) and 15\,{\textmu}L of TMSPMC (0.06\,mmol). The mixture was bath sonicated for 30\,s. Finally, 250\,{\textmu}L of concentrated ammonia solution was added dropwise under vigorous stirring. The reaction mixture was stirred at RT for 16\,h.
The mixture was transferred into a Falcon tube and centrifuged at 20,000\,{\texttimes}\,g for 20\,min (10\,\textdegree C). The supernatant collected from the previous centrifugation was further centrifuged at 40,000\,{\texttimes}\,g for 20\,min (10\,\textdegree C). The pellets from the initial and the supernatant centrifugation steps were combined and the tubes were refilled with methanol. These washing steps were repeated three times in total. After the final centrifugation, the supernatant was removed, and the pellet was resuspended in a small amount of methanol and stored at -20\,\textdegree C.

\subsubsection{Coating of silicated FNDs with azidated HPMA polymer}
HPMA was freshly recrystallized before use. Five-fold excess in mass of HPMA as needed for polymerization, was gently heated in a small amount of EtOAc until fully dissolved. The hot solution was filtered through a 0.22\,{\textmu}m PTFE filter. The filtered HPMA solution was gently heated and hexane was added until the solution becomes milky. The milky HPMA solution was placed in a freezer. The precipitate was collected by filtration through frit (S3) and dried under vacuum. Simultaneously, AIBN, a radical initiator, was gently heated in a small amount of ethanol to fully dissolve. Then, the solution was left for 30\,min at RT. Like HPMA, the AIBN solution was then placed in a freezer and processed in the same way.
For 6\,mg of silicated FNDs, a total of 375\,{\textmu}L of DMSO was used. 100\,{\textmu}L of that amount of DMSO was added into a 4\,mL vial and the same volume of silicated FNDs in methanol was added. Methanol was removed using rotary evaporation (25\,\textdegree C, 10\,mM Hg). This step was repeated to transfer all of 6\,mg of silicated FNDs into DMSO. In the meantime, 125\,mg of HMPA (0.87\,mMol) and 37.6\,mg of AIBN (0.23\,mmol) were weighted in a 4\,mL vial and dissolved in 275\,{\textmu}L of DMSO. This mixture was added to the FNDs in DMSO. Finally, 6.56\,mg of AzMA (0.046\,mmol) was added. The vial was purged with argon for 20\,min. The reaction mixture was secured by argon and stirred for 72\,h at 55\,\textdegree C under argon atmosphere.
The mixture was transferred into four Eppendorf tubes and centrifuged at 20,000\,{\texttimes}\,g for 20\,min (15\,\textdegree C). The supernatant collected from the previous centrifugation was further centrifuged at 40,000\,{\texttimes}\,g for 20\,min (15\,\textdegree C). The pellets from the initial and the supernatant centrifugation steps were combined and the tubes were refilled with methanol. These washing steps were repeated four times in total. After the final centrifugation, the supernatant was removed, and the pellet was resuspended in methanol to reach 10.0\,mg/mL concentration and stored at -20\,\textdegree C.

\subsubsection{DNA binding to azidated HPMA-FNDs}
The sequence used was 5'-DBCO-TEG-TTT (TTC)$_{7}$ T-3'. 200\,{\textmu}L of FND-HPMA solution (2\,mg of FNDs) was pipetted into a 4\,mL vial and 200\,{\textmu}L of 8\,\texttimes\,PBS was added dropwise to the FND solution. In the meantime, ssDNA was dissolved in RNase-free water to a final concentration of 0.3\,{\textmu}mol/ml. 200\,{\textmu}L of the prepared DNA solution (60\,nmol) was mixed with 200\,{\textmu}L of 20\,mM TRIS buffer (pH = 8). The DNA solution was added dropwise to the FNDs while stirring. The mixture was incubated at 45\,\textdegree C for 21 days with stirring.
After the reaction, 1\,mL of water preheated to 45\,\textdegree C was added in steps of 100\,{\textmu}L each to the vial. The reaction mixture was cooled down to 35\,\textdegree C. The reaction mixture was then transferred to two Eppendorf tubes and centrifuged at 20,000\,{\texttimes}\,g for 20\,min (20\,\textdegree C). The supernatant from the first centrifugation was further centrifuged at 55,000\,{\texttimes}\,g for 25\,min (20\,\textdegree C). The pellets from both centrifugations were combined. These washing steps were repeated two times with RNase-free water, three times with DMSO and twice again with RNase-free water. After the last centrifugation, the final volume was adjusted for concentration of 10\,mg/mL. Sample was stored in a freezer at -20\,\textdegree C.

\subsection{Origami-particle hybridization and purification}
To create sufficient assemblies in the desired configuration, the binding of FNDs and AuNPs is done separately. First, the FND solution is incubated at 50\,\textdegree C for 30\,min to dissolve the gel phase of DNA and polymer on the particle surface. The FNDs are then sonicated for 5\,min before mixing. Origami nanostructures and FNDs are mixed in a molar ratio of 1:2. The solution is finally brought to 1\,{\texttimes}\,TAE and 10\,mM MgCl$_2$ and incubated for around 2\,h, starting at 48\,\textdegree C (T$>$T$_M$ of the connecting strand between origami structure and FND) with a steady decrease to RT.
Right after incubation, the freshly purified AuNPs are added in a 1:1 molar ratio (in respect to the origami concentration) to the origami-FND mixture. If necessary, the MgCl$_2$ concentration is again adjusted to 10\,mM. The solution is incubated at RT for around 90\,min.
To separate excess AuNPs and unbound origami structures from the desired product, the samples are purified \textit{via} agarose gel electrophoresis in a 0.7\,\%\,w/v agarose gel containing 1\,{\texttimes}\,TAE, 10\,mM MgCl$_2$, and 1\,{\texttimes}\,SYBR\texttrademark Safe (Thermo Fisher Scientific, USA). The samples are mixed with loading buffer (1\,{\texttimes}\,TAE, 1\,mM MgCl$_2$, 30\,\%\,v/v glycerol) in a 1:5 volume ratio and loaded into the gel. The gel is run for 120\,min at 60\,V and the bands of interest are cut out. Recovery of the sample from the gel happens in two steps. First, the cut band is squeezed between two Parafilm-covered glass slides and any liquid is pipetted off of the surface. The residual agarose that still contains sample is then processed \textit{via} Freeze 'N Squeeze Spin Columns (Bio-Rad, USA).

\subsection{Sample evaluation}
Purified assemblies are evaluated by AFM on a Bruker ICON Dimension in ScanAsyst mode after being deposited on freshly cleaved mica (Nano-Tec Muscovite, Micro to Nano, Netherlands) and incubated for at least 10\,min. Three 10\,{\texttimes}\,10\,{\textmu}m scans for imaging yield evaluation are taken. The images are then processed using Gwyddion software. The imaging yield is collected by hand-annotating the AFM scans and evaluating them by a custom-made ImageJ macro.
For further evaluation of optical properties \textit{via} FLIM, a composite substrate is prepared. To be able to measure the same micron-sized area using both techniques subsequently, coverslips are labeled with chromium-based fiducial markers prepared \textit{via} optical lithography. Then, a thin flake of freshly cleaved mica is glued to the surface of the coverslip using 1\,{\textmu}L of Norland Optical Adhesive 88 (Norland Products, USA). The glue is cured by irradiating the modified coverslip with a 365\,nm UV lamp for around 20\,min. The sample is then immediately deposited on the mica surface in the same way as described before for pure AFM imaging. Either 20\,{\texttimes}\,20\,{\textmu}m or 40\,{\texttimes}\,40\,{\textmu}m high resolution scans of defined areas were prepared in the same AFM setup as described above.

For sample evaluation \textit{via} TEM, carbon-coated copper grids are ozone-treated for 5 min, followed by incubation with 5\,{\textmu}L of 0.5\,M MgCl$_2$ solution for 3\,min. The excess liquid is wicked off the grid with filter paper and 5\,{\textmu}L of sample solution is deposited on the grid for 5\,min. Excess solution is again wicked off the grid with filter paper. The sample is stained with 2\,wt.\% uranyl acetate solution. For that, 5\,{\textmu}L solution is deposited for 1\,min, wicked off, and reapplied for another 10\,s. The grids are left to dry. The measurements are performed on a Jeol JEM-2100Plus Ultra High-Resolution TEM at 200\,kV.

\subsection{Fluorescence Lifetime Imaging}
FLIM was conducted on a MicroTime200 confocal microscopy setup (PicoQuant, Germany) using a 532\,nm pulsed laser at 5\,MHz repetition rate. The inverted microscope is equipped with a water immersion objective (UPlanSApo 60\,{\texttimes}, NA 1.2, Olympus). The laser spot is focused on the sample surface and lifetime and intensity maps are acquired over an area of 50\,{\texttimes}\,50\,{\textmu}m. Instrument response functions (IRF) are recorded individually on each sample. The fluorescence signal is collected through the same objective. The excess laser light was filtered by a dichroic mirror (ZT405/532rpc, Chroma, USA) and the signal was guided through two optical filters (550LP and 650LP) onto a single-photon avalanche diode (\texttau -SPAD, PicoQuant, Germany)

An overlay of AFM and FLIM measurements is created manually and appropriate structures are identified for lifetime analysis (for details see Supplementary Note 4). Using SymPhoTime64 software (PicoQuant, Germany), the selected structures are thresholded through manual selection from the overall image and analyzed \textit{via} iterative reconvolution fitting of a three-exponential function.

\subsection{Fluorescence Spectroscopy}
Spectra were acquired in the same confocal setup as Fluorescence Lifetime Imaging using and Andor Shamrock 303i spectrometer with Andor iXon 888 camera (Oxford Instruments, United Kingdom). The laser spot is focused on individual structures of known configuration. The signal is recorded with an accumulation time of 10\,s with 50 accumulations.

\medskip
\section*{Supporting Information} \par 
See the Supplementary Information for detailed data on FND characterization methods and data, AFM sample analysis, and optical characterization data of assemblies, as well as DNA origami staples.

\medskip
\printbibliography

\medskip
\section*{Acknowledgements} \par
This work was supported by Czech Science Foundation, grant number 21-17847M (NH, VP) and Czech Academy of Sciences, Premie lumina quaeruntur LQ200402101 (DR, VP). PC acknowledges funding by the Czech Academy of Sciences - Strategy AV21 (VP29), Czech Science Foundation project no. 23-04876S, European Union project C-QuENS (grant no. 101135359), Horizon Europe MSCA-SE project FLORIN (grant agreement ID: 101086142), Technology Agency of the Czech Republic, project TH90010001 EXTRASENS (ERA-NET/QuantERA Cofund Project), and the Ministry of Education, Youth, and Sports of the Czech Republic (project No. CZ.02.01.01/00/22\_008/0004558, co-funded by the European Union). The authors acknowledge the assistance provided by the Research Infrastructure NanoEnviCz, supported by the Ministry of Education, Youth, and Sports of the Czech Republic under Project No. LM2023066. We thank Olga Pavlatov\'{a}, V\'{a}clav Protiva, Peter Kapusta, Luka Pirker, and Jakub Jungwirth for valuable help. We further thank Helge Ewers and Stephanie Reich for their help in the initial stages of this work.

\section*{Author Contribution}
NH designed, optimized, and evaluated the assembly of nanoparticles, performed FLIM measurement and subsequent analysis, FS and DR assisted with the FLIM measurement and analysis, JC and MK optimized, prepared, and characterized the DNA functionalized nanodiamonds, IB helped with the DNA origami design, VP and PC supervised the project (VP the DNA mediated assembly and PL modulation, PC the nanodiamond functionalization). All authors contributed to the interpretation and discussion of the result. NH and VP wrote the manuscript with the contribution of all authors.

\end{document}


\begin{center}
    {\Large Supplementary Information}
\end{center}

\begin{center}
    \textbf{\Large High-Yield Assembly of Plasmon-Coupled Nanodiamonds via DNA Origami for Tailored Emission}
\end{center}
\vspace{1ex}

\begin{center}
    {\large{}
        Niklas Hansen$^{1,3}$, 
        Jakub \v{C}op\'{a}k $^{2,4}$, 
        Marek Kindermann$^{2}$,
        David Roesel$^{1}$, 
        Federica Scollo$^{1}$, 
        Ilko Bald$^{1,6}$
        Petr C\'{i}gler$^{2*}$, 
        Vladim\'{i}ra Petr\'{a}kov\'{a}$^{1,5*}$}
\end{center}

\begin{center}
    {\footnotesize $^{1}$J. Heyrovsk\'{y} Institute of Physical Chemistry, Czech Academy of Sciences, Dolej\v{s}kova 3, 182 23 Prague, Czechia}\\
    {\footnotesize $^{2}$Institute of Chemistry and Biochemistry, Czech Academy of Sciences, Flemingovo nam\v{e}st\'{i} 542/2, 160 00 Prague, Czechia}\\
    {\footnotesize $^{3}$Department of Physical Chemistry, Faculty of Chemical Engineering, University of Chemistry and Technology,\\
    Technick\'{a} 5, 166 28 Prague 6, Czechia}\\
    {\footnotesize $^{4}$Department of Physical and Macromolecular Chemistry, Faculty of Science, Charles University,\\
    Hlavova 2030/8, 128 40 Prague 2, Czechia}\\
    {\footnotesize $^{5}$Faculty of Biomedical Engineering, Czech Technical University, Nam\v{e}st\'{i} S\'{i}tn\'{a} 3105, 27201 Kladno 2, Czechia}\\
    {\footnotesize $^{6}$Institut f\"{u}r Chemie, Universit\"{a}t Potsdam, Karl-Liebknecht-Stra{\ss}e 24-25, 14476 Potsdam, Germany}\\
\vspace{12pt}
    {\footnotesize $^{*}$Correspondence to: vladimira.petrakova@jh-inst.cas.cz, petr.cigler@uochb.cas.cz}
\end{center}

\vfill{}

 \begin{center}
        \begin{tabular}{|l|>{\raggedright}m{11.5cm}|}
        \hline 
        \textbf{Supplementary Note 1} & Fluorescent Nanodiamond Characterization \tabularnewline
        \hline         
        \hline 
        \textbf{Supplementary Note 2} & Imaging yield analysis of DNA origami-FND assemblies \tabularnewline
        \hline   
        \hline 
        \textbf{Supplementary Note 3} & Imaging yield analysis of DNA origami-FND-AuNP assemblies \tabularnewline
        \hline   
        \hline 
        \textbf{Supplementary Note 4} & Correlative Microscopy Measurements and Lifetime Analysis \tabularnewline
        \hline   
        \hline 
        \textbf{Supplementary Note 5} & 12-helix staple list and modifications \tabularnewline
        \hline 
        \end{tabular}
    \end{center}
    
\vfill{}

\clearpage

\normalsize

\thispagestyle{empty}

\section*{Supplementary Note 1: Fluorescent Nanodiamond Characterization}

\subsection*{Dynamic Light Scattering and Zeta Potential measurements}
Hydrodynamic diameter and zeta potential of the samples were measured at room temperature using a Zetasizer Nano ZS (Malvern Instruments, UK) equipped with disposable folded capillary cells (DTS1070). Samples were prepared at a concentration of 0.1\,mg/mL. For each measurement, three independent runs were conducted, and results are reported as the average and standard deviation (Std).\\

\begin{table}[ht]
\centering
\caption{Dynamic light scattering (DLS) and zeta potential (ZP) measurements of fluorescent nanodiamonds (FNDs) after each step of the surface functionalization process.
The hydrodynamic diameter increases progressively from 60\,nm for bare FNDs to 116\,nm after silica coating, poly(HPMA) grafting, and DNA conjugation, confirming a successful build up of the surface layer. Zeta potential values also changed characteristically, reflecting changes in surface chemistry. The most negative potential was observed after silica coating (-32.6\,mV), followed by a shift toward less negative values upon polymer functionalization and slightly more negative due to DNA functionalization.}
\begin{tabular}{lcccc}
                     & \textbf{DLS [nm]} & \textbf{Std} & \textbf{Zeta potential [mV]} & \textbf{Std}  \\ 
\toprule
Bare FND             & 60.2              & 0.6          & -23.1                        & 1.6           \\ 
\midrule
FND-silica             & 84.2              & 1.2          & -32.6                        & 0.5           \\ 
\midrule
FND-silica-pHPMA     & 97.3              & 1.2          & -19.1                        & 1.1           \\ 
\midrule
FND-silica-pHPMA-DNA & 115.9             & 0.5          & -21.7                        & 0.3           \\
\bottomrule
\end{tabular}
\label{tab:DLS-zeta}
\end{table}

\clearpage
\subsection*{X-ray Photon Spectroscopy Analysis}
Surface elemental composition and chemical states were investigated by X-ray photoelectron spectroscopy using a SPECS spectrometer with a monochromatic Al K{\textalpha} source (1486.7\,eV) and a hemispherical Phoibos 150 electron analyzer. Survey spectra were acquired at a pass energy (E$_p$) of 100\,eV, while high-resolution scans were recorded at E$_p$ = 30\,eV. All measurements were conducted under ultra-high vacuum (base pressure <10$^{-9}$\,mbar). To\,minimize surface charging, a low-energy electron flood gun (30\,{\textmu}A, 3.5\,eV) was employed throughout the measurements.\\

\begin{figure}[h]
    \includegraphics[width=0.7\textwidth]{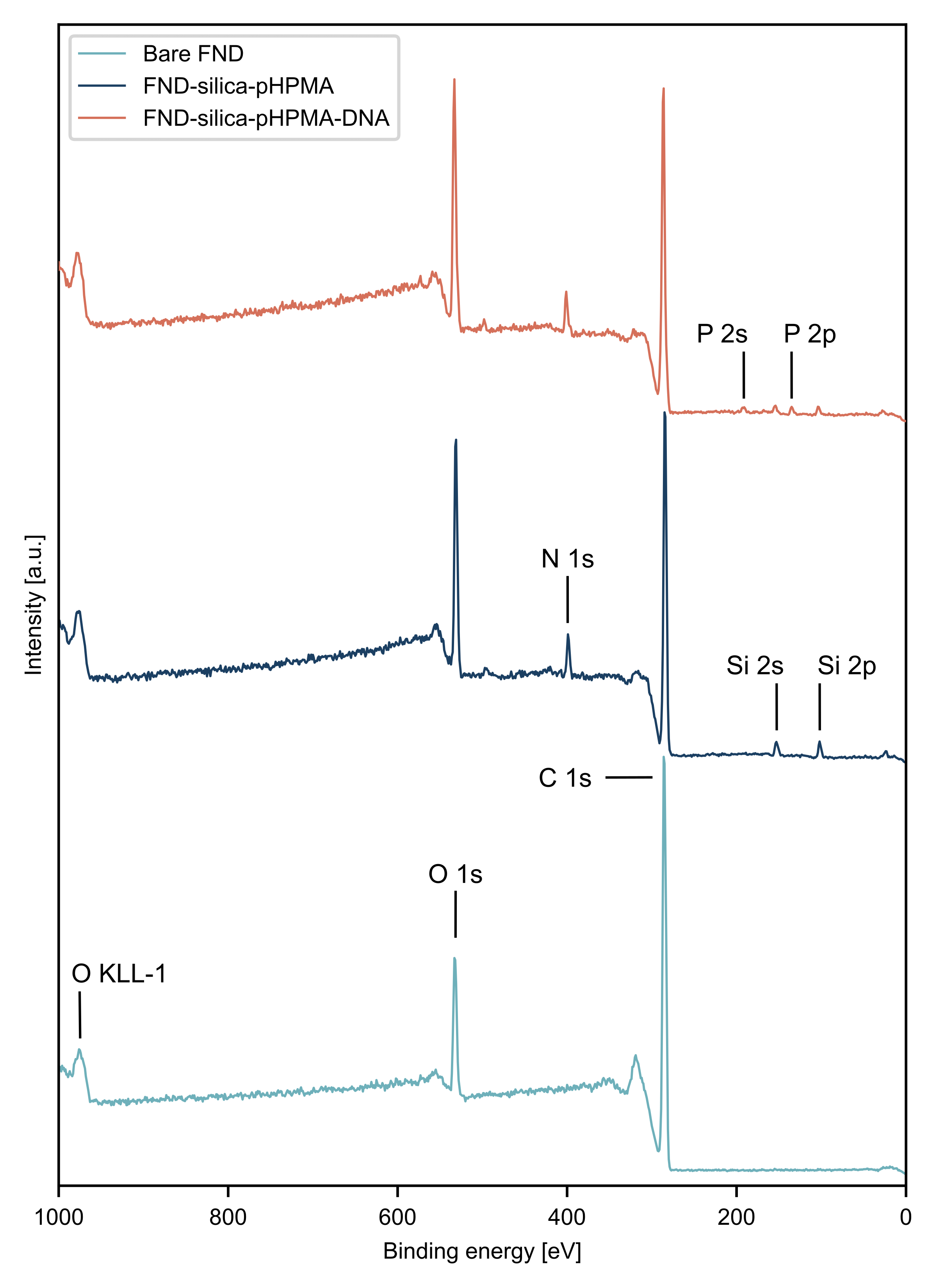}
    \centering
    \caption{X-ray photoelectron spectroscopy (XPS) survey spectra of bare fluorescent nanodiamonds (Bare FND, light blue), FNDs modified with silica layer and poly(HPMA) (FND-silica-pHPMA, dark blue), and DNA-functionalized FND-pHPMA (FND-silica-pHPMA-DNA, orange). 
The bare FNDs show characteristic C 1s (286.0 eV) and O 1s (533.0\,eV) peaks, with a carbon content of 89.7\,at\% and surface oxygen 10.3\,at\% due to the presence of surface oxygen-containing functional groups. These values were used as a baseline for comparison with subsequent surface modifications. Upon polymer grafting on silica layer (FND-silica-pHPMA), new peaks appeared for N 1s (399.0\,eV; 4.2\,at\%) and Si 2p (102 eV; 4.0\,at\%), along with increased O 1s intensity (17.2\,at\%) and decreased C 1s intensity (74.6\,at\%), indicating successful attachment of silica and the polymer coating. The appearance of N 1s (4.15\,at\%) is consistent with the presence of amide functionalities in the poly(HPMA) coating while Si 2s and 2p peaks reflects the presence of silica layer. After conjugation with DNA (FND-silica-pHPMA-DNA), an increase in N 1s (4.7\,at\%) and the emergence of P 2p and P 2s peaks (1.4\,at\%) correspond to the presence of DNA phosphate groups. The observed surface composition changes across samples confirm the stepwise functionalization of the nanodiamond surface.
}
    \label{fig:XPS}
\end{figure}

\clearpage
\subsection*{Thermogravimetric Analysis}
Thermal stability and compositional analysis of the samples were performed using a TGA 5500 instrument (TA Instruments, USA). Approximately 1\,mg of sample was placed in a platinum crucible and subjected to a heating ramp of 2\,\textdegree /min from 25\,\textdegree to 900\,\textdegree. Prior to heating, the sample was held isothermally at 25\,\textdegree for 20\,min to ensure thermal equilibration. The measurements were carried out under a constant inert gas flow of 35\,mL/min.\\

\begin{figure}[h]
    \includegraphics[width=.8\textwidth]{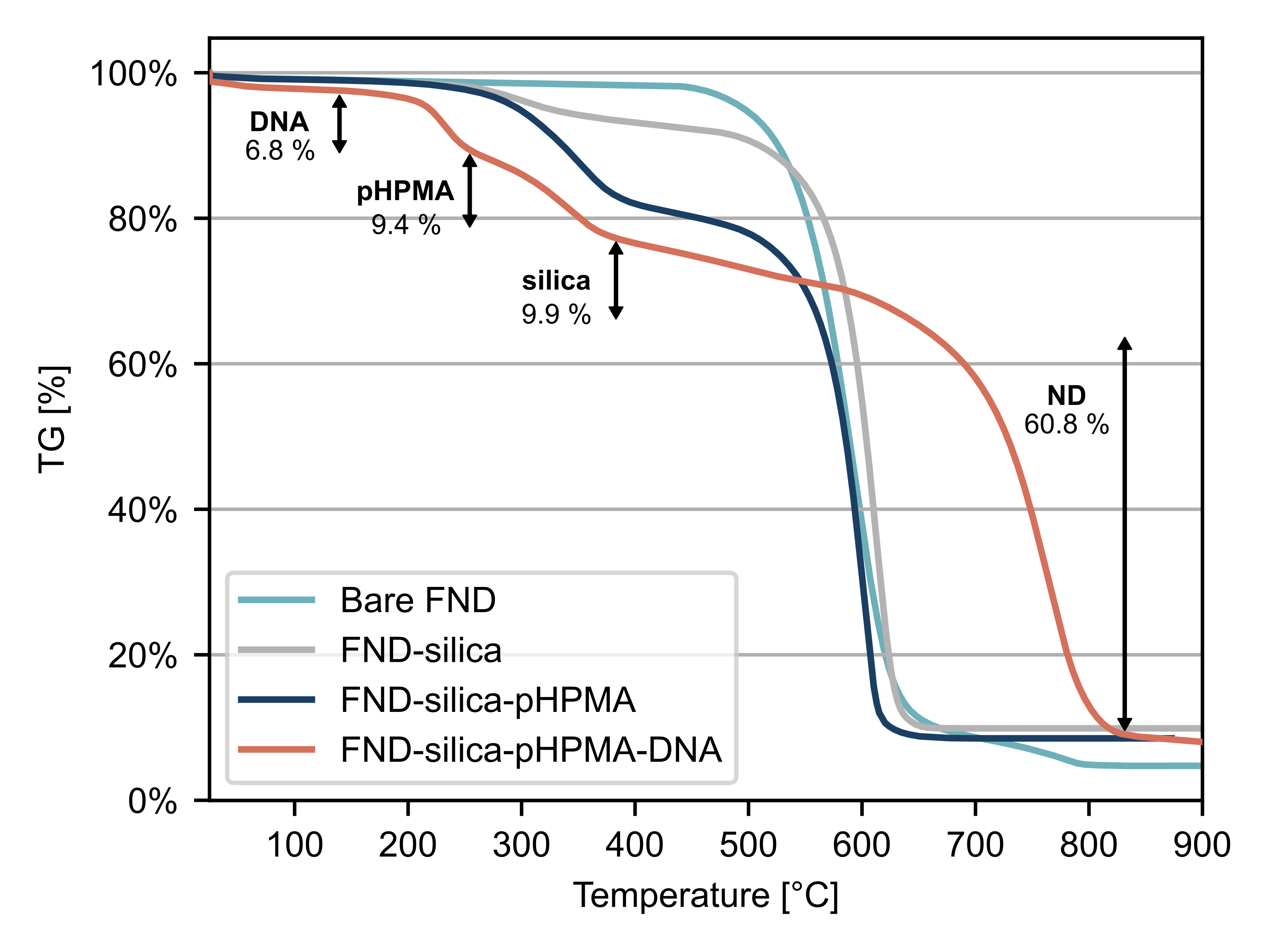}
    \centering
    \caption{Thermogravimetric analysis (TGA) of fluorescent nanodiamonds (FNDs) at various stages of surface modification.
Curves are shown for bare FNDs (light blue), silica-coated FNDs (FND-silica, grey), poly(HPMA)-functionalized FND-silica (FND-silica-pHPMA, dark blue), and DNA-conjugated particles (FND-silica-pHPMA-DNA, orange). Bare FNDs exhibited high thermal stability, with a major mass loss occurring around 525\,\textdegree C, corresponding to the oxidative degradation of the nanodiamond core. Upon silica coating, a gradual weight loss between 250 and 550\,\textdegree C was observed, consistent with decomposition of the organic components of the silica layer. Subsequent grafting with poly(HPMA) introduced an additional mass loss starting at approximately 250\,\textdegree C, attributed to thermal degradation of the polymer. Finally, the introduction of DNA led to a distinct weight loss onset at approximately 175\,\textdegree C, corresponding to the decomposition of the oligonucleotide chains. Notably, the combustion temperature of nanodiamonds increased by approximately 100\,\textdegree C in the presence of DNA, as confirmed by measurements with a simple FND/DNA mixture (data not shown). The sequential mass loss steps observed in the orange curve confirm the successful and stepwise surface functionalization of the nanodiamonds.
}
    \label{fig:TGA}
\end{figure}

\clearpage
\subsection*{Nanoparticle Tracking Analysis}
Particle concentrations were determined using a NanoSight NS300 system (Malvern Panalytical, UK) equipped with a 405\,nm violet laser (maximum output <70\,mW). Sample was diluted to reach an optimal concentration range corresponding to 50\,-\,70 particles per frame. Eight videos of 180\,s each were recorded and analyzed using NanoSight software. Reported values represent the mean of all eight measurements.\\

\begin{figure}[h]
    \includegraphics[width=.8\textwidth]{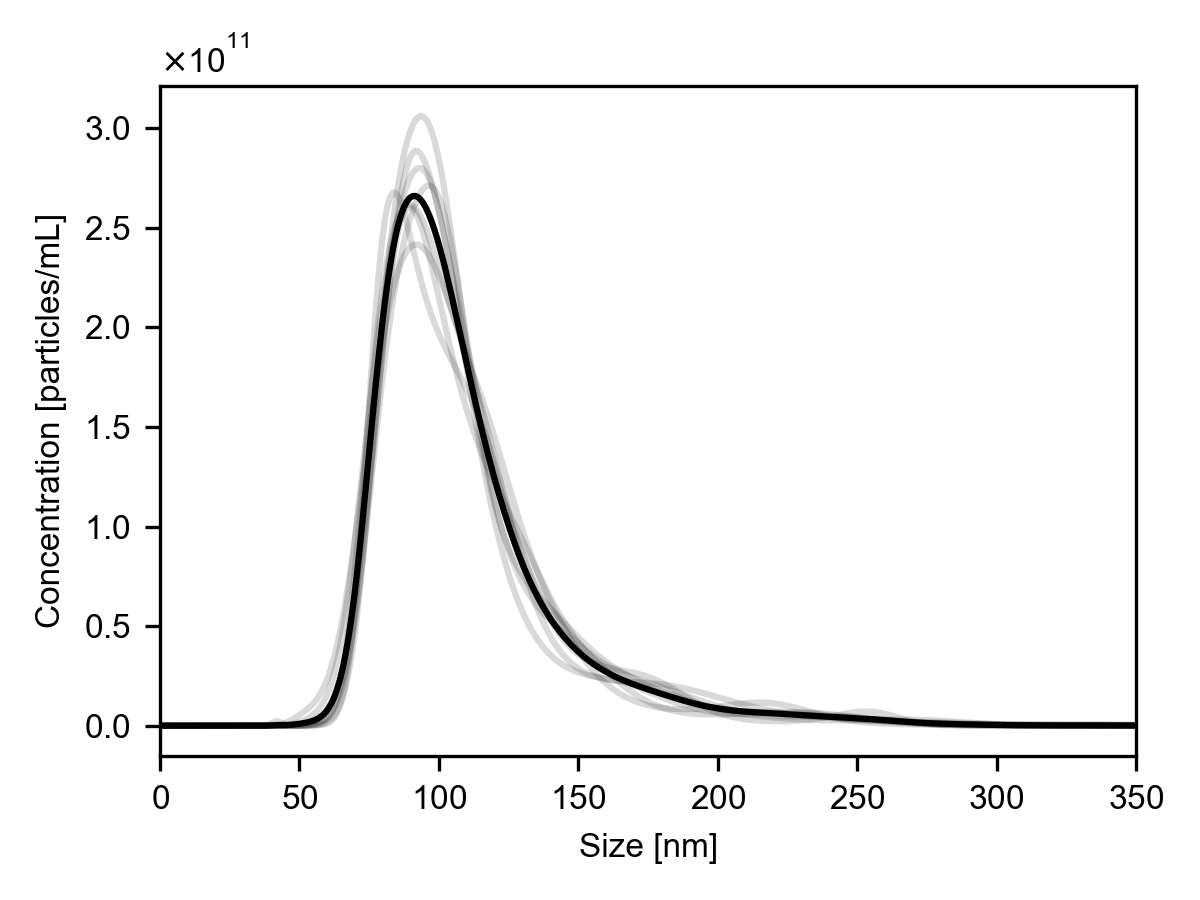}
    \centering
    \caption{Nanoparticle Tracking Analysis (NTA) of DNA-conjugated particles (FND-silica-pHPMA-DNA). The size distribution profile shows the particle concentration as a function of size. All measurements were performed at a dilution factor of 1\,:\,20,000. The grey curves represent individual measurements (n\,=\,8), while the black curve corresponds to the averaged value. A dominant single peak at 90–100\,nm (average value 92\,nm) is observed across all replicates, consistent with well-dispersed DNA-coated nanodiamond particles. The narrow distribution and zero aggregation indicate high colloidal stability maintained throughout the surface modification steps.
}
    \label{fig:NTA}
\end{figure}

\clearpage
\subsection*{Determination of DNA Loading of FNDs}
Fluorescence measurements were carried out using a Tecan Spark multimode microplate reader (Software version 2.3) with Thermo Scientific\texttrademark Nunc 384-Well Optical Bottom Plates. Samples were measured at a concentration of 0.1\,mg/mL in a volume of 25\,{\textmu}L per well. The excitation wavelength was set to 462\,nm, and emission spectra were recorded in the range of 507\,-\,600\,nm. Fluorescence intensity at 526\,nm (the emission maximum) was used to construct the calibration curve.\\

\begin{figure}[h]
    \includegraphics[width=.8\textwidth]{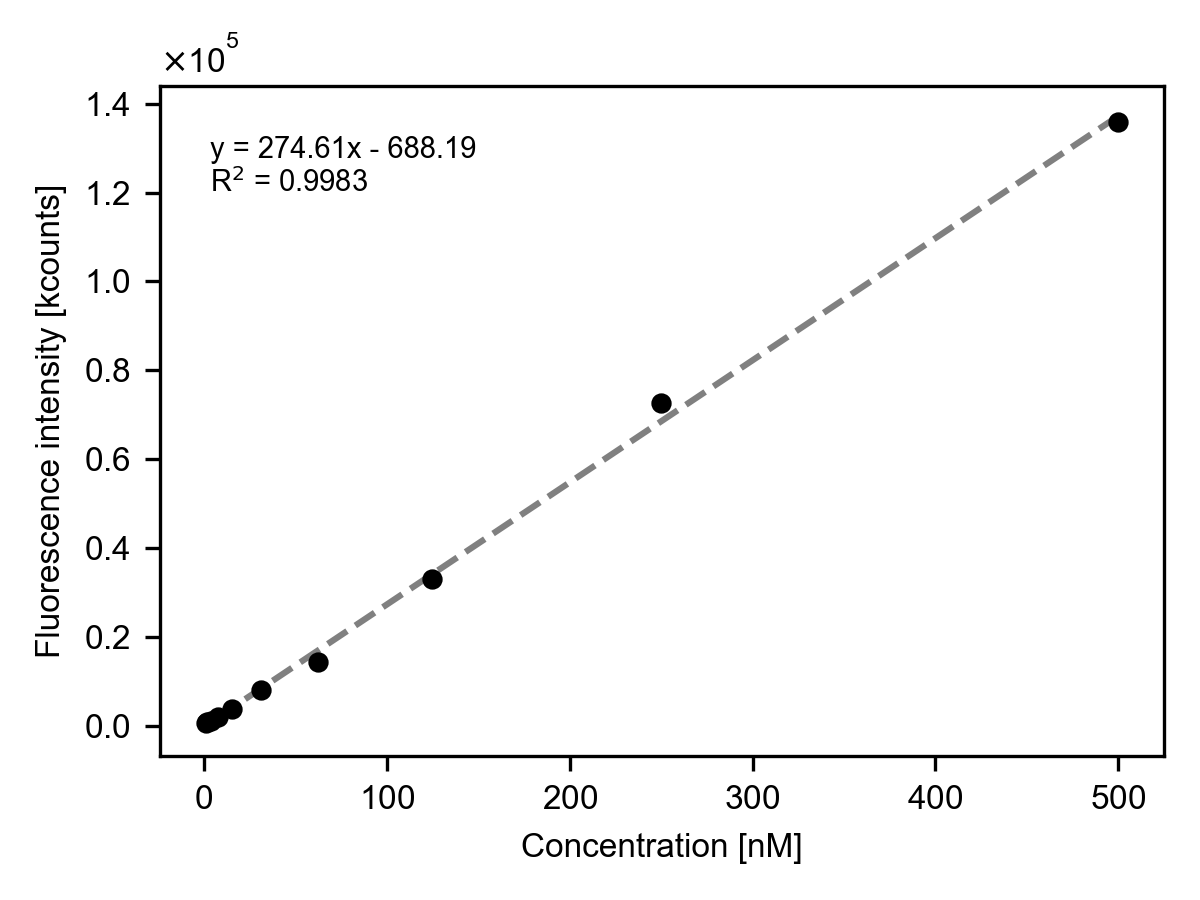}
    \centering
    \caption{Fluorescence calibration curve for determination of surface bound ssDNA on FNDs}
    \label{fig:ssDNA}
\end{figure}

\noindent To determine the number of DNA strands conjugated per ND, a hybridization assay with complementary DNA labeled at the 5' end with ATTO488 dye (cDNA-ATTO488) was employed. The oligonucleotide (5'-/5ATTO488N/ AGA AGA AGA AGA AGA AGA AGA AAA A-3') was purchased from IDT with HPLC purification.

\subsubsection*{Hybridization with cDNA-ATTO488}
A 50\,µl of NDs-HMPA-DNA colloidal solution (10\,mg/ml; 0.5\,mg NDs total) was mixed with 374\,µl RNase-free water in a 1.5\,ml Eppendorf tube. To this solution, 5\,µl of 1\,M tris(hydroxymethyl)aminomethane (TRIS; pH = 8.0), 6.25\,µl of 1\,M magnesium chloride (MgCl$_2$), and 65\,µl of 0.1\,mM solution cDNA-ATTO488 were sequentially added. The mixture was briefly vortexed and then incubated in a 95\,\textdegree C water bath. After 10\,min, the mixture was allowed to slowly cool to RT over 18 hours.
After the reaction, the reaction mixture was centrifuged at 20,000\,{\texttimes}\,g for 20\,min (4\,\textdegree C). The supernatant from the first centrifugation was further centrifuged at 55,000\,{\texttimes}\,g for 20\,min (4\,\textdegree C). The pellets from both centrifugations were combined together. These washing steps were repeated four times with RNase-free water. After last centrifugation, the final volume was adjusted for concentration of 1.00\,mg/ml.

\subsubsection*{Dehybridization of cDNA-ATTO488}
All dehybridization steps were performed in triplicate. A 100\,µl of the hybridized sample (1\,mg/ml; 0.1\,mg ND in total) was mixed with 900\,µl of 10mM (4-(2-hydroxyethyl)-1-piperazineethanesulfonic acid) (HEPES; pH = 7.4) prepared in RNase-free water. The solution was incubated at 95\,\textdegree C for 10\,min, then immediately centrifuged in a pre-heated rotor at 55,000\,{\texttimes}\,g for 20\,min (40\,\textdegree C). The supernatant, containing released cDNA-ATTO488, was collected for fluorescence quantification. The pellet was resuspended by bath sonication and adjusted to concentration of 1\,mg/ml.
Fluorescence Quantification and Data Analysis
A calibration curve was prepared using cDNA-ATTO488 standards (0.98 – 500\,nM) in 10\,mM HEPES buffer (pH = 7.4) containing NDs-HMPA at a final concentration of 0.1\,mg/ml (see Figure S\ref{fig:ssDNA}). Fluorescence intensity at 526\,nm (emission maximum) was used for quantification. Based on the calibration equation, the concentration of released cDNA-ATTO488 (ccDNA-ATTO488) in the dehybridization supernatant was calculated using:

\begin{equation*}
    c_{cDNA-ATTO488}=  \frac{(F_{supernanat}+intercept)}{slope} = \frac{98 045 + 688.19}{274.61} = 360\,nM
\end{equation*}
\\
\noindent The concentration of cDNA-ATTO488 was converted to the amount of substance using the following equation in a volume of one milliliter:

\begin{equation*}
    n_{cDNA-ATTO488} = (c_{cDNA-ATTO488}\times10^{-9})\times(1\times10^{-3})  = 
    3.60\times10^{-10}\,mol
\end{equation*}
\\
\noindent And using Avogadro's constant to calculate the number of molecules:

\begin{equation*}
N_{cDNA-ATTO488} = n_{cDNA-ATTO488} \times 6.022\times10^{23} = 2.17\times10^{14}
\end{equation*}
\\
\noindent The number of nanodiamonds in one milliliter in the corresponding pellet was determined by NTA, yielding:

\begin{equation*}
N_{ND from NTA} = 1.11\times10^{12}
\end{equation*}
\\
\noindent Therefore, the average number of DNA strands per ND was calculated as:

\begin{equation*}
N_{DNA/ND} = \frac{N_{cDNA-ATTO488}}{N_{ND from NTA}} =\frac{2.17\times10^{14}}{1.11\times10^{12}} = 195
\end{equation*}
\\
\noindent The final result is reported as the mean ± standard error from three independent measurements:

\begin{equation*}
N_{DNA/ND} = 198\pm6
\end{equation*}

\clearpage

\section*{Supplementary Note 2: Imaging yield analysis of DNA origami-FND assemblies}

\begin{figure}[ht]
    \centering
    \includegraphics{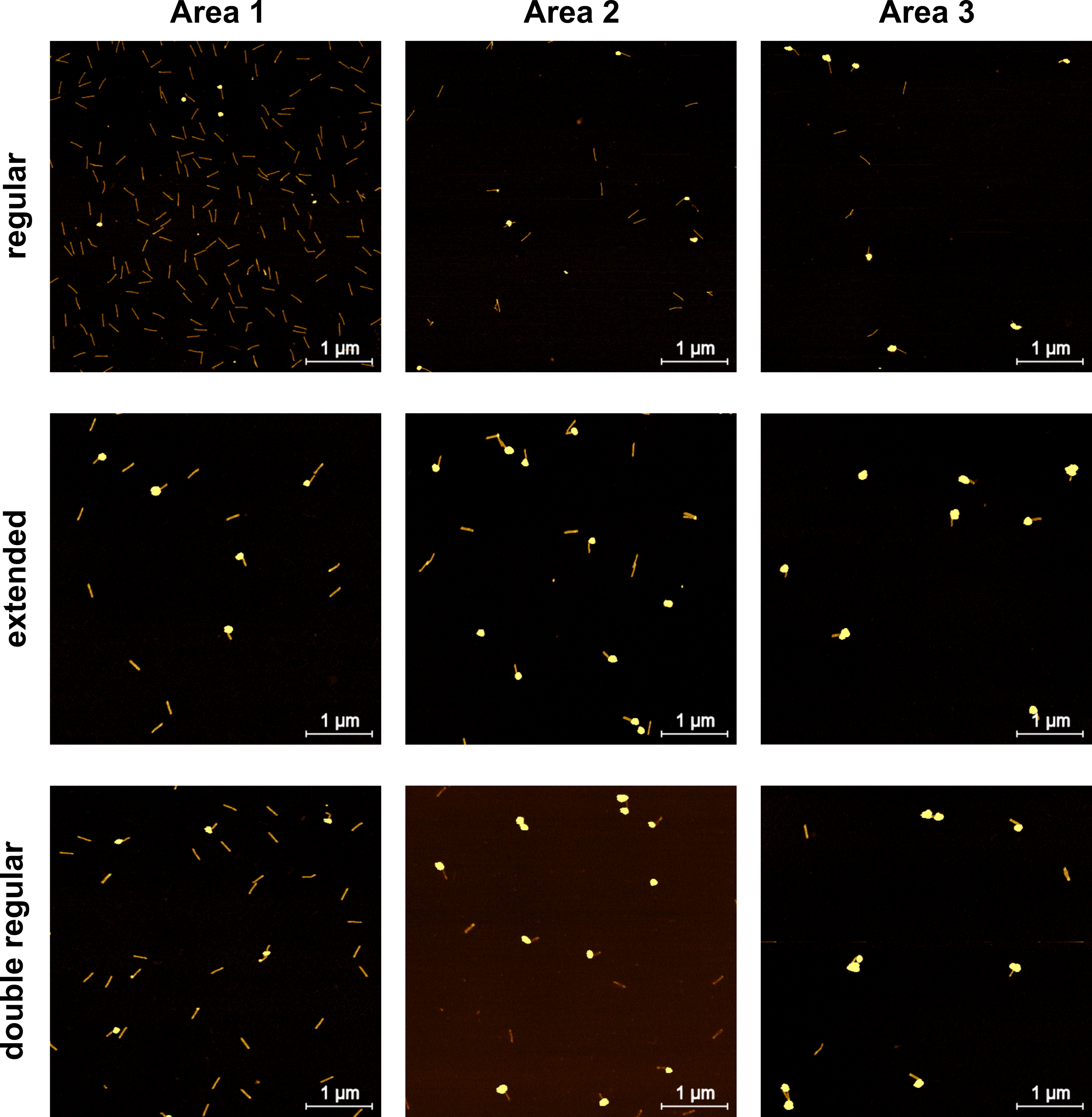}
    \caption{Exemplary images of different sample configurations for binding only FNDs to 12HB structures. The "Area" designation marks the extracted and imaged area in the gel lane.}
    \label{fig:SI_only-FND_AFM}
\end{figure}

In this section, we list the data of the AFM analysis corresponding to assemblies containing only fluorescent nanodiamonds on 12-helix bundles in different configurations. The analysis is described in the Method section. Example AFM images for each sample are shown in Figure S\ref{fig:SI_only-FND_AFM}.\\

\begin{table}[ht]
\caption{Overview of the evaluated number of structures for binding of FNDs to 12HB nanostructures and their corresponding percentage from the total number of evaluated structures.}
\centering
\begin{tabular}{llccccc}
\begin{sideways}\end{sideways}                &                         & \textbf{bare 12HB} & \textbf{12HB-FND} & \textbf{excluded} & \textbf{Total} &   \\ 
\cmidrule[\heavyrulewidth]{2-6}
\multirow{6}{*}{\rotcell{\begin{tabular}[c]{@{}l@{}}\hspace{-2em}\textbf{regular}\end{tabular}}} & \multirow{2}{*}{Area 1} & 792                & 9                 & 42                & 843            &   \\
                                              &                         & 94.0\%             & 1.1\%             & 5.0\%             &                &   \\ 
\cmidrule{2-6}
                                              & \multirow{2}{*}{Area 2} & 120                & 55                & 66                & 241            &   \\
                                              &                         & 49.8\%             & 22.8\%            & 27.4\%            &                &   \\ 
\cmidrule{2-6}
                                              & \multirow{2}{*}{Area 3} & 68                 & 45                & 23                & 136            &   \\
                                              &                         & 50.0\%             & \textbf{33.1\%}            & 16.9\%            &                &   \\ 
\cmidrule[\heavyrulewidth]{2-6}
\begin{sideways}\end{sideways}                &                         &                    &                   &                   &                &   \\
\multirow{6}{*}{\rotcell{\begin{tabular}[c]{@{}l@{}}\hspace{-2em}\textbf{extended}\end{tabular}}}  & \multirow{2}{*}{Area 1} & 88                 & 24                & 7                 & 119            &   \\
                                              &                         & 73.9\%             & 20.2\%            & 5.9\%             &                &   \\ 
\cmidrule{2-6}
                                              & \multirow{2}{*}{Area 2} & 84                 & 102               & 36                & 222            &   \\
                                              &                         & 37.8\%             & 45.9\%            & 16.2\%            &                &   \\ 
\cmidrule{2-6}
                                              & \multirow{2}{*}{Area 3} & 43                 & 100               & 29                & 172            &   \\
                                              &                         & 25.0\%             & \textbf{58.1\%}            & 16.9\%            &                &   \\
\cmidrule[\heavyrulewidth]{2-6}
\end{tabular}
\vspace{0.8cm}

\begin{tabular}{llccccc}
                                                   &                         & \textbf{bare 12HB} & \textbf{12HB-FND (1x)} & \textbf{12HB-FND (2x)} & \textbf{excluded} & \textbf{Total}  \\ 
\cmidrule[\heavyrulewidth]{2-7}
\multirow{6}{*}{\rotcell{\begin{tabular}[c]{@{}l@{}}\hspace{-3em}\textbf{double regular}\end{tabular}}} & \multirow{2}{*}{Area 1} & 1064               & 90                     & 4                      & 81                & 1239            \\
                                                   &                         & 85.9\%             & 7.3\%                  & 0.3\%                  & 6.5\%             &                 \\ 
\cmidrule{2-7}
                                                   & \multirow{2}{*}{Area 2} & 197                & 166                    & 13                     & 62                & 438             \\
                                                   &                         & 45.0\%             & 37.9\%                 & 3.0\%                  & 14.2\%            &                 \\ 
\cmidrule{2-7}
                                                   & \multirow{2}{*}{Area 3} & 41                 & 85                     & 10                     & 40                & 176             \\
                                                   &                         & 23.3\%             & \textbf{48.3\%}                 & \textbf{5.7\%}                  & 22.7\%            &                 \\
\cmidrule[\heavyrulewidth]{2-7}
\end{tabular}
\end{table}

\clearpage

\clearpage

\section*{Supplementary Note 3: Imaging yield analysis of DNA origami-FND-AuNP assemblies}

\begin{figure}[ht]

    \centering
    \includegraphics[width=.9\textwidth]{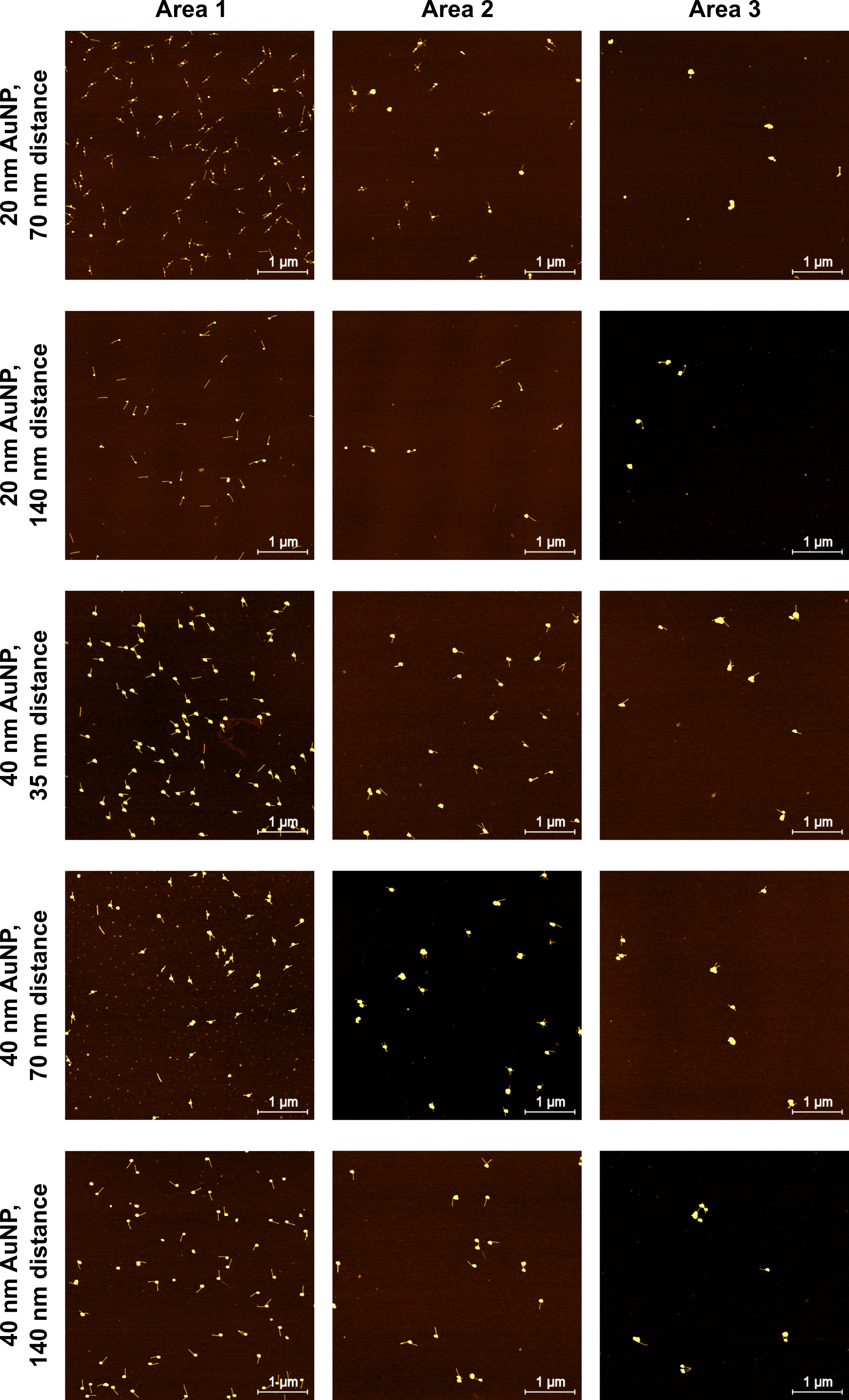}
    \caption{Exemplary images of different sample configurations for binding FNDs and AuNPs of different size to 12HB structures with "regular" binding sites. The "Area" designation marks the extracted and imaged area in the gel lane.}
    \label{fig:SI_AFM_reg}
\end{figure}

\begin{figure}[ht]
    \centering
    \includegraphics{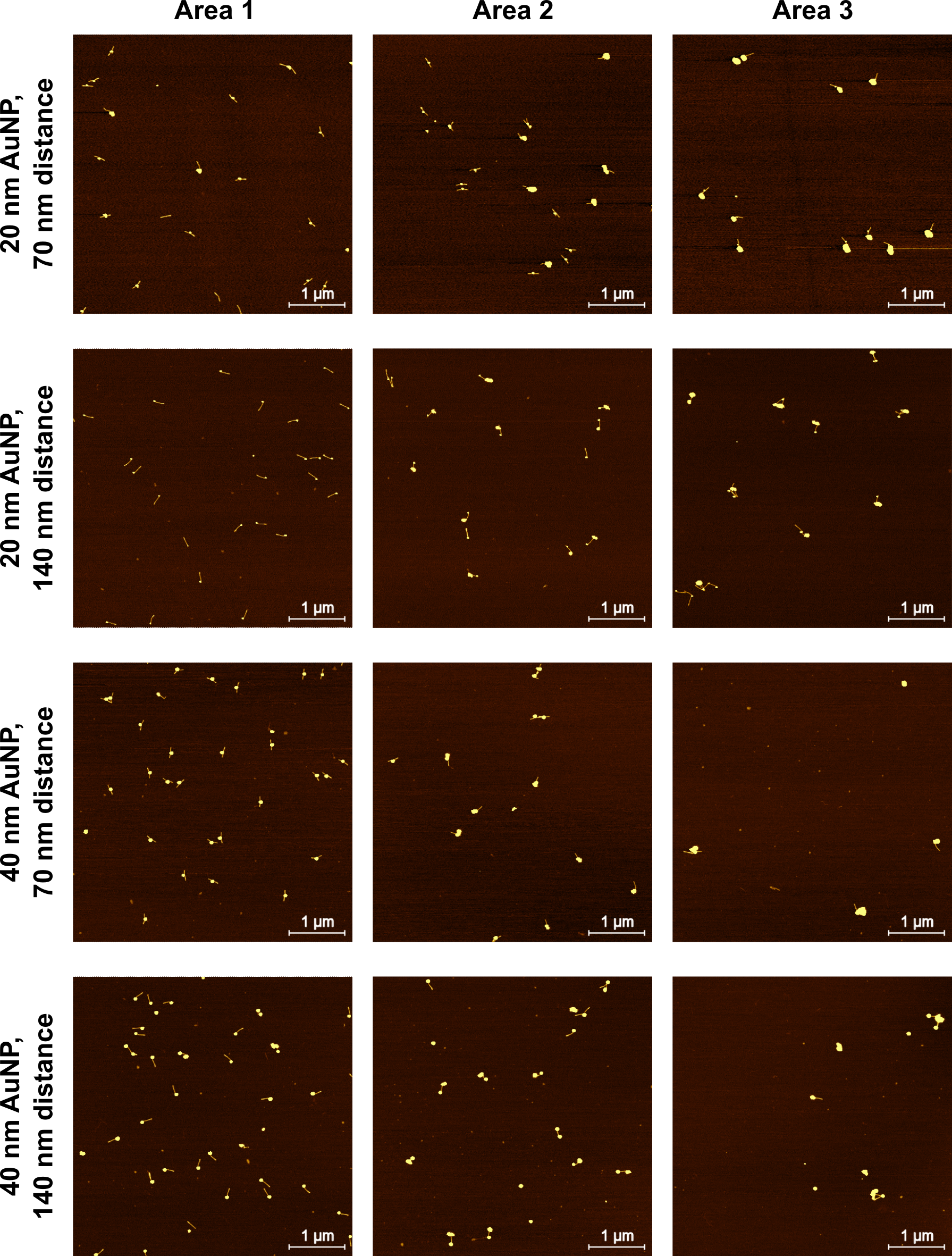}
    \caption{Exemplary images of different sample configurations for binding FNDs and AuNPs of different size to 12HB structures with the "extended" FND binding site. The "Area" designation marks the extracted and imaged area in the gel lane.}
    \label{fig:SI_AFM_ext}
\end{figure}

\begin{figure}[ht]
    \centering
    \includegraphics{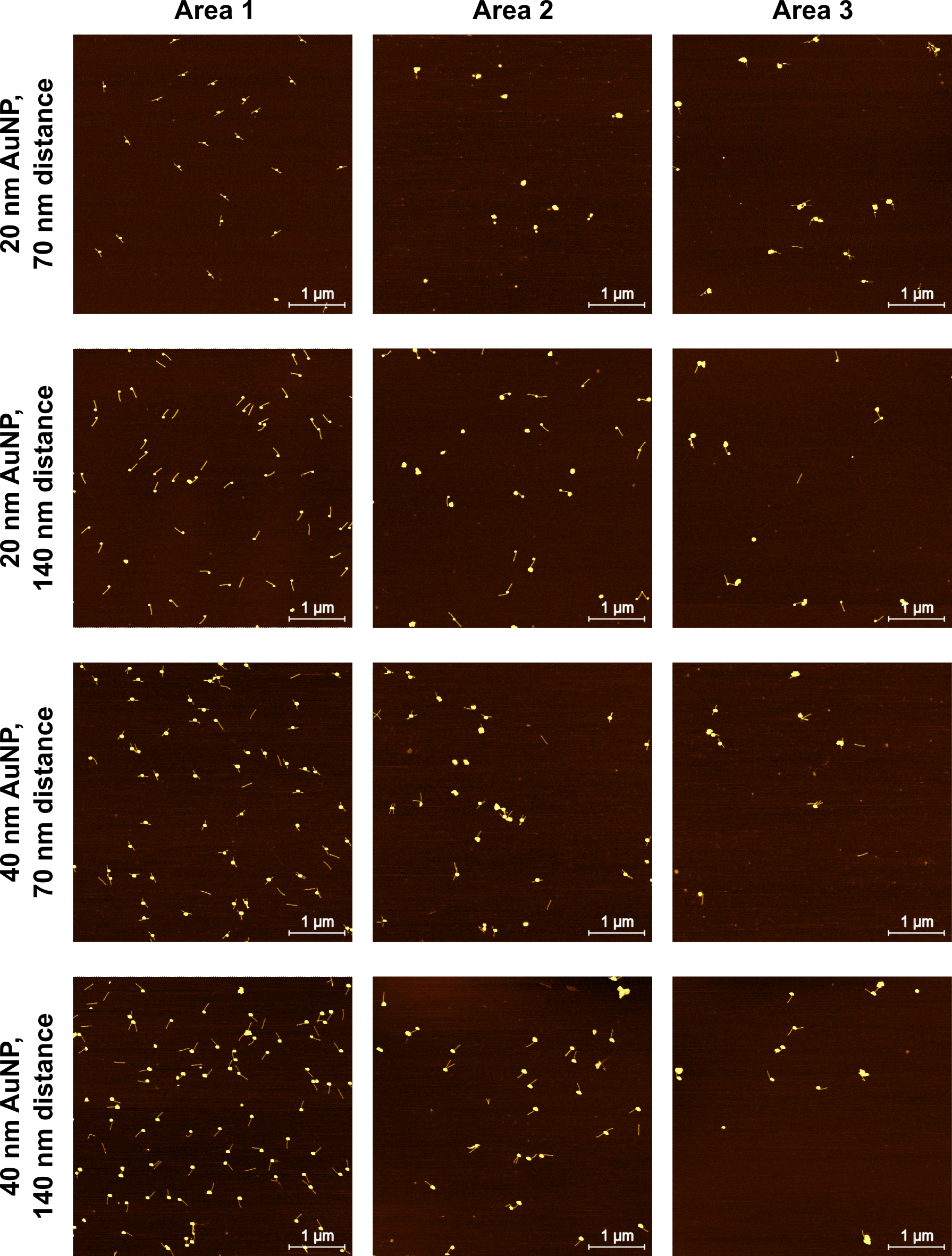}
    \caption{Exemplary images of different sample configurations for binding FNDs and AuNPs of different size to 12HB structures with the "shifted" AuNP binding site. The "Area" designation marks the extracted and imaged area in the gel lane.}
    \label{fig:SI_AFM_shift}
\end{figure}

Here, we list the data of the AFM analysis corresponding to assemblies containing fluorescent nanodiamonds and different gold nanoparticles on 12-helix bundles in different configurations. The analysis is described in the Method section. Example AFM images for each sample are shown in Figure S\ref{fig:SI_AFM_reg} ("regular" configuration), Figure S\ref{fig:SI_AFM_ext} ("extended" configuration, and Figure S\ref{fig:SI_AFM_shift} ("shifted" configuration).\\

 \small \justifying \noindent \textbf{Table S 3:} Overview of the evaluated number of structures for different AuNP size and interparticle distance configurations for the "regular" binding site option and their corresponding percentage from the total number of evaluated structures.

\begin{longtable}{llcccccc}
\begin{sideways}\end{sideways}                                                                                     &                         & \textbf{bare 12HB} & \textbf{12HB-AuNP} & \textbf{12HB-FND} & \textbf{assembly} & \textbf{excluded} & \textbf{Total}  \\ 
\cmidrule[\heavyrulewidth]{2-8}
\multirow{6}{*}{\rotcell{\begin{tabular}[c]{@{}l@{}}\hspace{-3.5em}\textbf{20 nm AuNP, }\\\hspace{-3.5em}\textbf{70 nm distance}\end{tabular}}}  & \multirow{2}{*}{Area 1} & 16                 & 297                & 4                 & 6                 & 24                & 347             \\
                                                                                                                   &                         & 4.6\%              & 85.6\%             & 1.2\%             & 1.7\%             & 6.9\%             &                 \\ 
\cmidrule{2-8}
                                                                                                                   & \multirow{2}{*}{Area 2} & 2                  & 52                 & 13                & 43                & 100               & 210             \\
                                                                                                                   &                         & 1.0\%              & 24.8\%             & 6.2\%             & \textbf{20.5\%}            & 47.6\%            &                 \\ 
\cmidrule{2-8}
                                                                                                                   & \multirow{2}{*}{Area 3} & 1                  & 2                  & 9                 & 10                & 29                & 51              \\
                                                                                                                   &                         & 2.0\%              & 3.9\%              & 17.6\%            & 19.6\%            & 56.9\%            &                 \\ 
\cmidrule[\heavyrulewidth]{2-8}
\begin{sideways}\end{sideways}                                                                                     &                         &                    &                    &                   &                   &                   &                 \\
\multirow{6}{*}{\rotcell{\begin{tabular}[c]{@{}l@{}}\hspace{-3.5em}\textbf{20 nm AuNP, }\\\hspace{-3.5em}\textbf{140 nm distance}\end{tabular}}} & \multirow{2}{*}{Area 1} & 41                 & 224                & 20                & 11                & 24                & 320             \\
                                                                                                                   &                         & 12.8\%             & 70.0\%             & 6.3\%             & 3.4\%             & 7.5\%             &                 \\ 
\cmidrule{2-8}
                                                                                                                   & \multirow{2}{*}{Area 2} & 4                  & 17                 & 9                 & 25                & 47                & 102             \\
                                                                                                                   &                         & 3.9\%              & 16.7\%             & 8.8\%             & 24.5\%            & 46.1\%            &                 \\ 
\cmidrule{2-8}
                                                                                                                   & \multirow{2}{*}{Area 3} & 8                  & 6                  & 2                 & 28                & 20                & 64              \\
                                                                                                                   &                         & 12.5\%             & 9.4\%              & 3.1\%             & \textbf{43.8\%}            & 31.3\%            &                 \\ 
\cmidrule[\heavyrulewidth]{2-8}
\begin{sideways}\end{sideways}                                                                                     &                         &                    &                    &                   &                   &                   &                 \\
\multirow{6}{*}{\rotcell{\begin{tabular}[c]{@{}l@{}}\hspace{-3.5em}\textbf{40 nm AuNP,}\\\hspace{-3.5em}\textbf{35 nm distance}\end{tabular}}}  & \multirow{2}{*}{Area 1} & 40                 & 471                & 25                & 1                 & 22                & 559             \\
                                                                                                                   &                         & 7.2\%              & 84.3\%             & 4.5\%             & 0.2\%             & 3.9\%             &                 \\ 
\cmidrule{2-8}
                                                                                                                   & \multirow{2}{*}{Area 2} & 5                  & 124                & 19                & 14                & 40                & 202             \\
                                                                                                                   &                         & 2.5\%              & 61.4\%             & 9.4\%             & 6.9\%             & 19.8\%            &                 \\ 
\cmidrule{2-8}
                                                                                                                   & \multirow{2}{*}{Area 3} & 7                  & 24                 & 10                & 13                & 37                & 91              \\
                                                                                                                   &                         & 7.7\%              & 26.4\%             & 11.0\%            & \textbf{14.3\%}            & 40.7\%            &                 \\ 
\cmidrule[\heavyrulewidth]{2-8}
\begin{sideways}\end{sideways}                                                                                     &                         &                    &                    &                   &                   &                   &                 \\
\multirow{6}{*}{\rotcell{\begin{tabular}[c]{@{}l@{}}\hspace{-3.5em}\textbf{40 nm AuNP,}\\\hspace{-3.5em}\textbf{70 nm distance}\end{tabular}}}  & \multirow{2}{*}{Area 1} & 41                 & 446                & 28                & 11                & 64                & 590             \\
                                                                                                                   &                         & 6.9\%              & 75.6\%             & 4.7\%             & 1.9\%             & 10.8\%            &                 \\ 
\cmidrule{2-8}
                                                                                                                   & \multirow{2}{*}{Area 2} & 3                  & 87                 & 21                & 20                & 64                & 195             \\
                                                                                                                   &                         & 1.5\%              & 44.6\%             & 10.8\%            & 10.3\%            & 32.8\%            &                 \\ 
\cmidrule{2-8}
                                                                                                                   & \multirow{2}{*}{Area 3} & 1                  & 13                 & 8                 & 7                 & 30                & 59              \\
                                                                                                                   &                         & 1.7\%              & 22.0\%             & 13.6\%            & \textbf{11.9\%}            & 50.8\%            &                 \\ 
\cmidrule[\heavyrulewidth]{2-8}
\begin{sideways}\end{sideways}                                                                                     &                         &                    &                    &                   &                   &                   &                 \\
\multirow{6}{*}{\rotcell{\begin{tabular}[c]{@{}l@{}}\hspace{-3.5em}\textbf{40 nm AuNP,}\\\hspace{-3.5em}\textbf{140 nm distance}\end{tabular}}} & \multirow{2}{*}{Area 1} & 20                 & 341                & 20                & 11                & 36                & 428             \\
                                                                                                                   &                         & 4.7\%              & 79.7\%             & 4.7\%             & 2.6\%             & 8.4\%             &                 \\ 
\cmidrule{2-8}
                                                                                                                   & \multirow{2}{*}{Area 2} & 2                  & 77                 & 22                & 21                & 51                & 173             \\
                                                                                                                   &                         & 1.2\%              & 44.5\%             & 12.7\%            & \textbf{12.1\%}            & 29.5\%            &                 \\ 
\cmidrule{2-8}
                                                                                                                   & \multirow{2}{*}{Area 3} & 4                  & 6                  & 7                 & 2                 & 8                 & 27              \\
                                                                                                                   &                         & 14.8\%             & 22.2\%             & 25.9\%            & 7.4\%             & 29.6\%            &                 \\ 
\cmidrule[\heavyrulewidth]{2-8}
\begin{sideways}\end{sideways}                                                                                     &                         &                    &                    &                   &                   &                   &                
\end{longtable}

\clearpage

\begin{table}[ht]
\caption{Overview of the evaluated number of structures for different AuNP size and interparticle distance configurations for the "extended" binding site option. Percentages highlighted in bold depict gel-optimized values visualized in Figure 3.}
\begin{tabular}{llcccccc}
                                                                                                                   &                         & \textbf{bare 12HB} & \textbf{12HB-AuNP} & \textbf{12HB-FND} & \textbf{assembly} & \textbf{excluded} & \textbf{Total}  \\ 
\cmidrule[\heavyrulewidth]{2-8}
\multirow{6}{*}{\rotcell{\begin{tabular}[c]{@{}l@{}}\hspace{-3.5em}\textbf{20 nm AuNP, }\\\hspace{-3.5em}\textbf{70 nm distance}\end{tabular}}}  & \multirow{2}{*}{Area 1} & 27                 & 140                & 11                & 3                 & 35                & 216             \\
                                                                                                                   &                         & 12.5\%             & 64.8\%             & 5.1\%             & 1.4\%             & 16.2\%            &                 \\ 
\cmidrule{2-8}
                                                                                                                   & \multirow{2}{*}{Area 2} & 11                 & 54                 & 21                & 28                & 41                & 155             \\
                                                                                                                   &                         & 7.1\%              & 34.8\%             & 13.5\%            & 18.1\%            & 26.5\%            &                 \\ 
\cmidrule{2-8}
                                                                                                                   & \multirow{2}{*}{Area 3} & 7                  & 5                  & 19                & 60                & 17                & 108             \\
                                                                                                                   &                         & 6.5\%              & 4.6\%              & 17.6\%            & \textbf{55.6\%}            & 15.7\%            &                 \\ 
\cmidrule[\heavyrulewidth]{2-8}
                                                                                                                   &                         &                    &                    &                   &                   &                   &                 \\
\multirow{6}{*}{\rotcell{\begin{tabular}[c]{@{}l@{}}\hspace{-3.5em}\textbf{20 nm AuNP, }\\\hspace{-3.5em}\textbf{140 nm distance}\end{tabular}}} & \multirow{2}{*}{Area 1} & 1                  & 277                & 8                 & 1                 & 10                & 297             \\
                                                                                                                   &                         & 0.3\%              & 93.3\%             & 2.7\%             & 0.3\%             & 3.4\%             &                 \\ 
\cmidrule{2-8}
                                                                                                                   & \multirow{2}{*}{Area 2} & 1                  & 24                 & 3                 & 63                & 22                & 113             \\
                                                                                                                   &                         & 0.9\%              & 21.2\%             & 2.7\%             & \textbf{55.8\%}            & 19.5\%            &                 \\ 
\cmidrule{2-8}
                                                                                                                   & \multirow{2}{*}{Area 3} & 1                  & 2                  & 4                 & 43                & 29                & 79              \\
                                                                                                                   &                         & 1.3\%              & 2.5\%              & 5.1\%             & 54.4\%            & 36.7\%            &                 \\ 
\cmidrule[\heavyrulewidth]{2-8}
                                                                                                                   &                         &                    &                    &                   &                   &                   &                 \\
\multirow{6}{*}{\rotcell{\begin{tabular}[c]{@{}l@{}}\hspace{-3.5em}\textbf{40 nm AuNP, }\\\hspace{-3.5em}\textbf{70 nm distance}\end{tabular}}}  & \multirow{2}{*}{Area 1} & 3                  & 271                & 3                 & 16                & 15                & 308             \\
                                                                                                                   &                         & 1.0\%              & 88.0\%             & 1.0\%             & 5.2\%             & 4.9\%             &                 \\ 
\cmidrule{2-8}
                                                                                                                   & \multirow{2}{*}{Area 2} & 0                  & 11                 & 4                 & 39                & 28                & 82              \\
                                                                                                                   &                         & 0.0\%              & 13.4\%             & 4.9\%             & \textbf{47.6\%}            & 34.1\%            &                 \\ 
\cmidrule{2-8}
                                                                                                                   & \multirow{2}{*}{Area 3} & 0                  & 3                  & 4                 & 6                 & 15                & 28              \\
                                                                                                                   &                         & 0.0\%              & 10.7\%             & 14.3\%            & 21.4\%            & 53.6\%            &                 \\ 
\cmidrule[\heavyrulewidth]{2-8}
                                                                                                                   &                         &                    &                    &                   &                   &                   &                 \\
\multirow{6}{*}{\rotcell{\begin{tabular}[c]{@{}l@{}}\hspace{-3.5em}\textbf{40 nm AuNP, }\\\hspace{-3.5em}\textbf{140 nm distance}\end{tabular}}} & \multirow{2}{*}{Area 1} & 10                 & 248                & 7                 & 19                & 19                & 303             \\
                                                                                                                   &                         & 3.3\%              & 81.8\%             & 2.3\%             & 6.3\%             & 6.3\%             &                 \\ 
\cmidrule{2-8}
                                                                                                                   & \multirow{2}{*}{Area 2} & 1                  & 18                 & 3                 & 49                & 15                & 86              \\
                                                                                                                   &                         & 1.2\%              & 20.9\%             & 3.5\%             & \textbf{57.0\%}            & 17.4\%            &                 \\ 
\cmidrule{2-8}
                                                                                                                   & \multirow{2}{*}{Area 3} & 2                  & 3                  & 2                 & 3                 & 8                 & 18              \\
                                                                                                                   &                         & 11.1\%             & 16.7\%             & 11.1\%            & 16.7\%            & 44.4\%            &                 \\ 
\cmidrule[\heavyrulewidth]{2-8}
                                                                                                                   &                         &                    &                    &                   &                   &                   &                
\end{tabular}
\end{table}

\clearpage

\clearpage

\begin{table}[ht]
\centering
\caption{Overview of the evaluated number of structures for different AuNP size and interparticle distance configurations for the "shifted" binding site option. Values highlighted in bold letters represent gel-optimized values shown in Figure 3.}
\begin{tabular}{llcccccc}
                                                                                                                   &                         & \textbf{bare 12HB} & \textbf{12HB-AuNP} & \textbf{12HB-FND} & \textbf{assembly} & \textbf{excluded} & \textbf{Total}  \\ 
\cmidrule[\heavyrulewidth]{2-8}
\multirow{6}{*}{\rotcell{\begin{tabular}[c]{@{}l@{}}\hspace{-3.5em}\textbf{20 nm AuNP, }\\\hspace{-3.5em}\textbf{70 nm distance}\end{tabular}}}  & \multirow{2}{*}{Area 1} & 17                 & 263                & 6                 & 6                 & 9                 & 301             \\
                                                                                                                   &                         & 5.6\%              & 87.4\%             & 2.0\%             & 2.0\%             & 3.0\%             &                 \\ 
\cmidrule{2-8}
                                                                                                                   & \multirow{2}{*}{Area 2} & 4                  & 17                 & 10                & 36                & 9                 & 76              \\
                                                                                                                   &                         & 5.3\%              & 22.4\%             & 13.2\%            & \textbf{47.4\%}            & 11.8\%            &                 \\ 
\cmidrule{2-8}
                                                                                                                   & \multirow{2}{*}{Area 3} & 2                  & 9                  & 14                & 44                & 29                & 98              \\
                                                                                                                   &                         & 2.0\%              & 9.2\%              & 14.3\%            & 44.9\%            & 29.6\%            &                 \\ 
\cmidrule[\heavyrulewidth]{2-8}
                                                                                                                   &                         &                    &                    &                   &                   &                   &                 \\
\multirow{6}{*}{\rotcell{\begin{tabular}[c]{@{}l@{}}\hspace{-3.5em}\textbf{20 nm AuNP, }\\\hspace{-3.5em}\textbf{140 nm distance}\end{tabular}}} & \multirow{2}{*}{Area 1} & 46                 & 376                & 6                 & 8                 & 25                & 461             \\
                                                                                                                   &                         & 10.0\%             & 81.6\%             & 1.3\%             & 1.7\%             & 5.4\%             &                 \\ 
\cmidrule{2-8}
                                                                                                                   & \multirow{2}{*}{Area 2} & 3                  & 97                 & 8                 & 51                & 52                & 211             \\
                                                                                                                   &                         & 1.4\%              & 46.0\%             & 3.8\%             & 24.2\%            & 24.6\%            &                 \\ 
\cmidrule{2-8}
                                                                                                                   & \multirow{2}{*}{Area 3} & 3                  & 5                  & 7                 & 14                & 22                & 51              \\
                                                                                                                   &                         & 5.9\%              & 9.8\%              & 13.7\%            & \textbf{27.5\%}            & 43.1\%            &                 \\ 
\cmidrule[\heavyrulewidth]{2-8}
                                                                                                                   &                         &                    &                    &                   &                   &                   &                 \\
\multirow{6}{*}{\rotcell{\begin{tabular}[c]{@{}l@{}}\hspace{-3.5em}\textbf{40 nm AuNP, }\\\hspace{-3.5em}\textbf{70 nm distance}\end{tabular}}}  & \multirow{2}{*}{Area 1} & 121                & 503                & 40                & 6                 & 59                & 729             \\
                                                                                                                   &                         & 16.6\%             & 69.0\%             & 5.5\%             & 0.8\%             & 8.1\%             &                 \\ 
\cmidrule{2-8}
                                                                                                                   & \multirow{2}{*}{Area 2} & 36                 & 149                & 40                & 27                & 81                & 333             \\
                                                                                                                   &                         & 10.8\%             & 44.7\%             & 12.0\%            & 8.1\%             & 24.3\%            &                 \\ 
\cmidrule{2-8}
                                                                                                                   & \multirow{2}{*}{Area 3} & 12                 & 21                 & 11                & 8                 & 25                & 77              \\
                                                                                                                   &                         & 15.6\%             & 27.3\%             & 14.3\%            & \textbf{10.4\%}            & 32.5\%            &                 \\ 
\cmidrule[\heavyrulewidth]{2-8}
                                                                                                                   &                         &                    &                    &                   &                   &                   &                 \\
\multirow{6}{*}{\rotcell{\begin{tabular}[c]{@{}l@{}}\hspace{-3.5em}\textbf{40 nm AuNP, }\\\hspace{-3.5em}\textbf{140 nm distance}\end{tabular}}} & \multirow{2}{*}{Area 1} & 74                 & 313                & 41                & 17                & 59                & 504             \\
                                                                                                                   &                         & 14.7\%             & 62.1\%             & 8.1\%             & 3.4\%             & 11.7\%            &                 \\ 
\cmidrule{2-8}
                                                                                                                   & \multirow{2}{*}{Area 2} & 17                 & 123                & 21                & 15                & 45                & 221             \\
                                                                                                                   &                         & 7.7\%              & 55.7\%             & 9.5\%             & 6.8\%             & 20.4\%            &                 \\ 
\cmidrule{2-8}
                                                                                                                   & \multirow{2}{*}{Area 3} & 13                 & 25                 & 29                & 30                & 34                & 131             \\
                                                                                                                   &                         & 9.9\%              & 19.1\%             & 22.1\%            & \textbf{22.9\%}            & 26.0\%            &                 \\ 
\cmidrule[\heavyrulewidth]{2-8}
                                                                                                                   &                         &                    &                    &                   &                   &                   &                
\end{tabular}
\end{table}

\clearpage

\clearpage

\clearpage

\section*{Supplementary Note 4: Correlative Microscopy Measurements and Lifetime Analysis}
\normalsize
Here, we present a detailed overview of the colocalized AFM and FLIM images, showing accurate matches between the height profile data of the AFM and the fluorescence signal recorded in FLIM. We highlight the analyzed structures and present the resulting fluorescence decay. This way of analyzing single structures has the advantage of reduced background influence on the signal and precise selection of the evaluated structures.

\begin{figure}[ht]
    \centering
    \includegraphics[width=.8\linewidth]{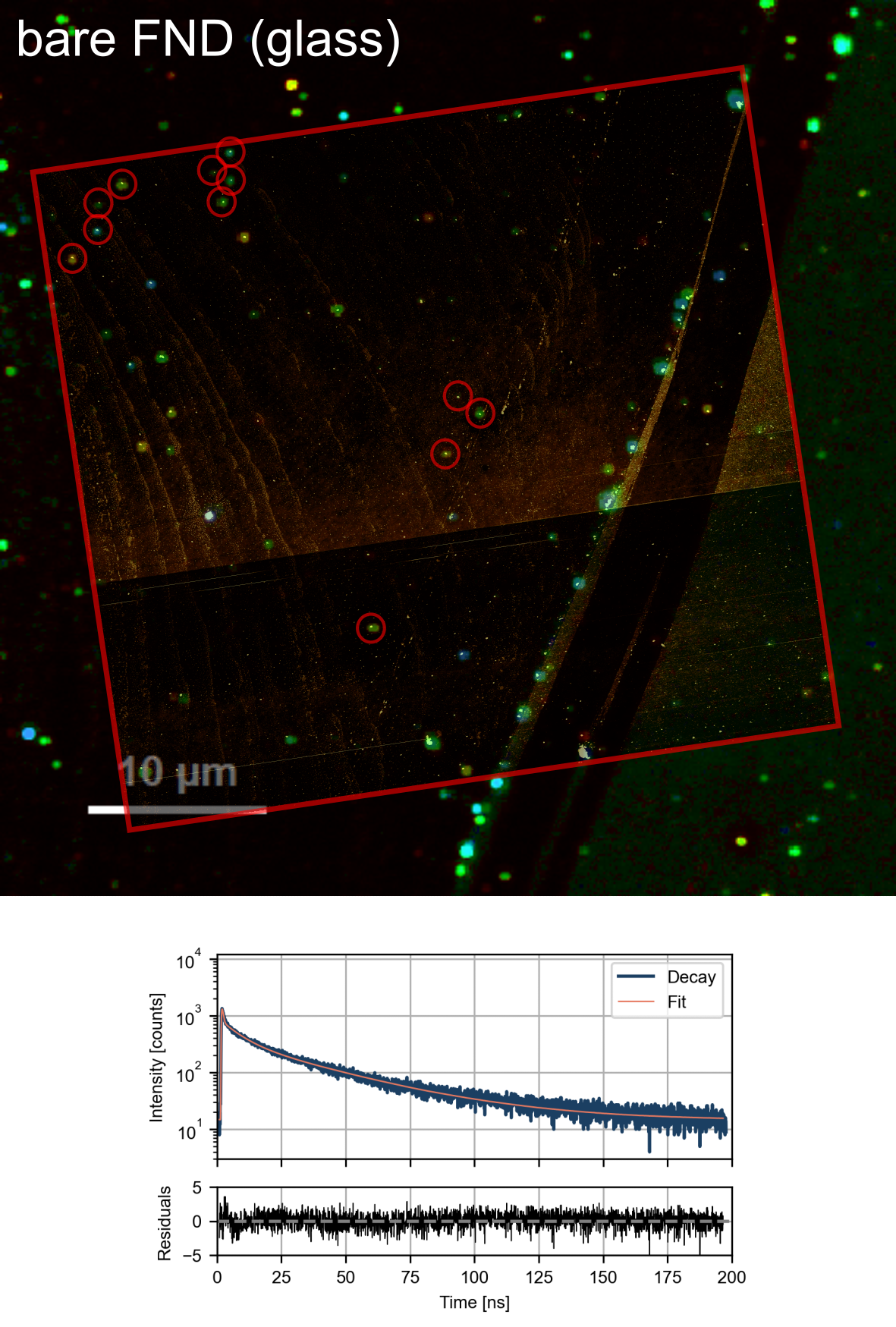}
    \caption{Overlay of high-resolution AFM image with FLIM measurement for bare fluorescent nanodiamonds on glass. Red circles highlight selected nanodiamonds. The graph shows the resulting fluorescence decay of the selected FNDs and the fit after analysis. Residuals represent goodness of fit with $\chi^2$ = 1.06.}
    \label{fig:overlay_FND-glass}
\end{figure}

\begin{figure}[ht]
    \centering
    \includegraphics[width=.8\linewidth]{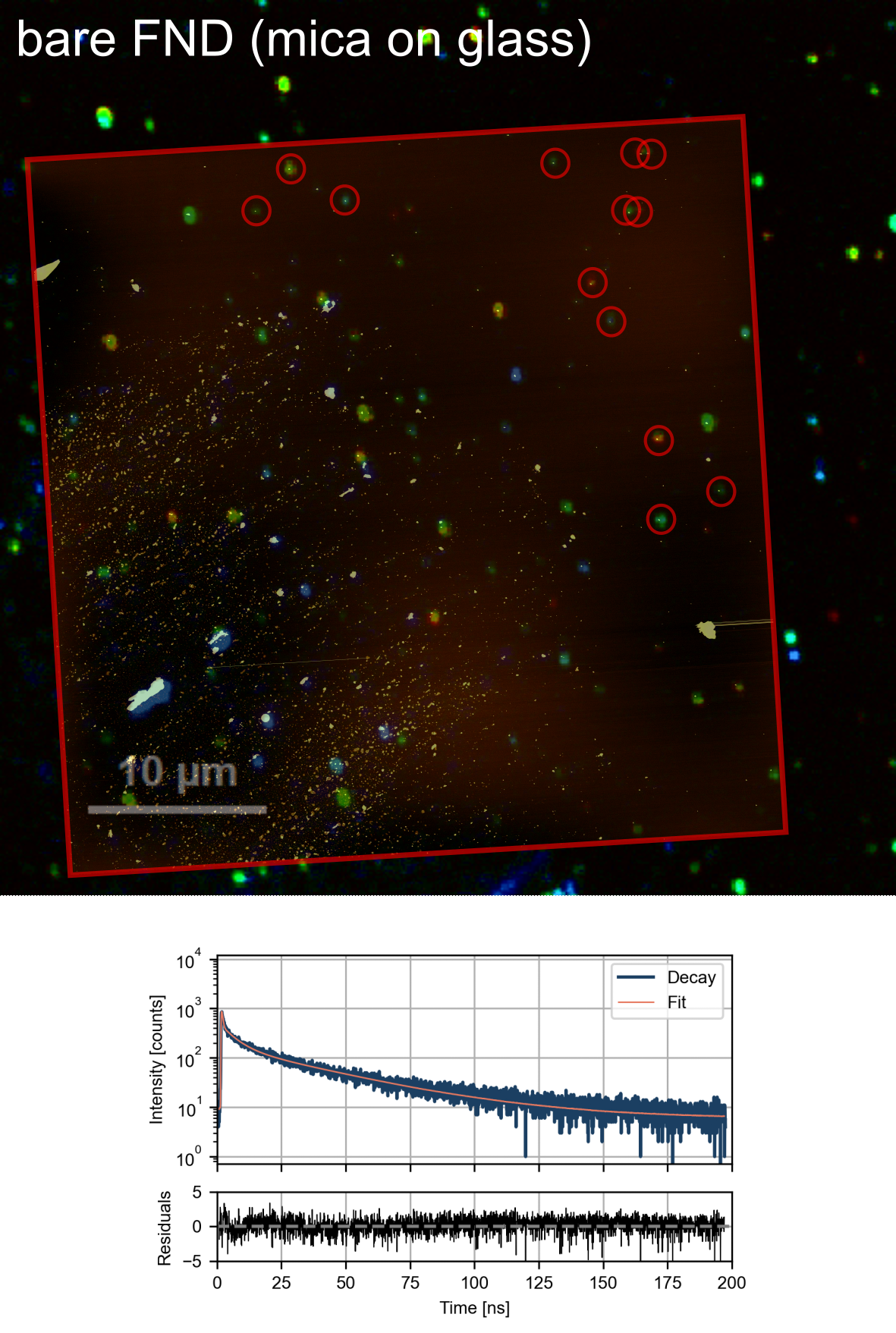}
    \caption{Overlay of high-resolution AFM image with FLIM measurement for bare fluorescent nanodiamonds on composite substrate consisting of mica on glass. Red circles highlight selected nanodiamonds. The graph shows the resulting fluorescence decay of the selected FNDs and the fit after analysis. Residuals represent goodness of fit with $\chi^2$ = 1.11}
    \label{fig:overlay_FND-mog}
\end{figure}

\begin{figure}[ht]
    \centering
    \includegraphics[width=.8\linewidth]{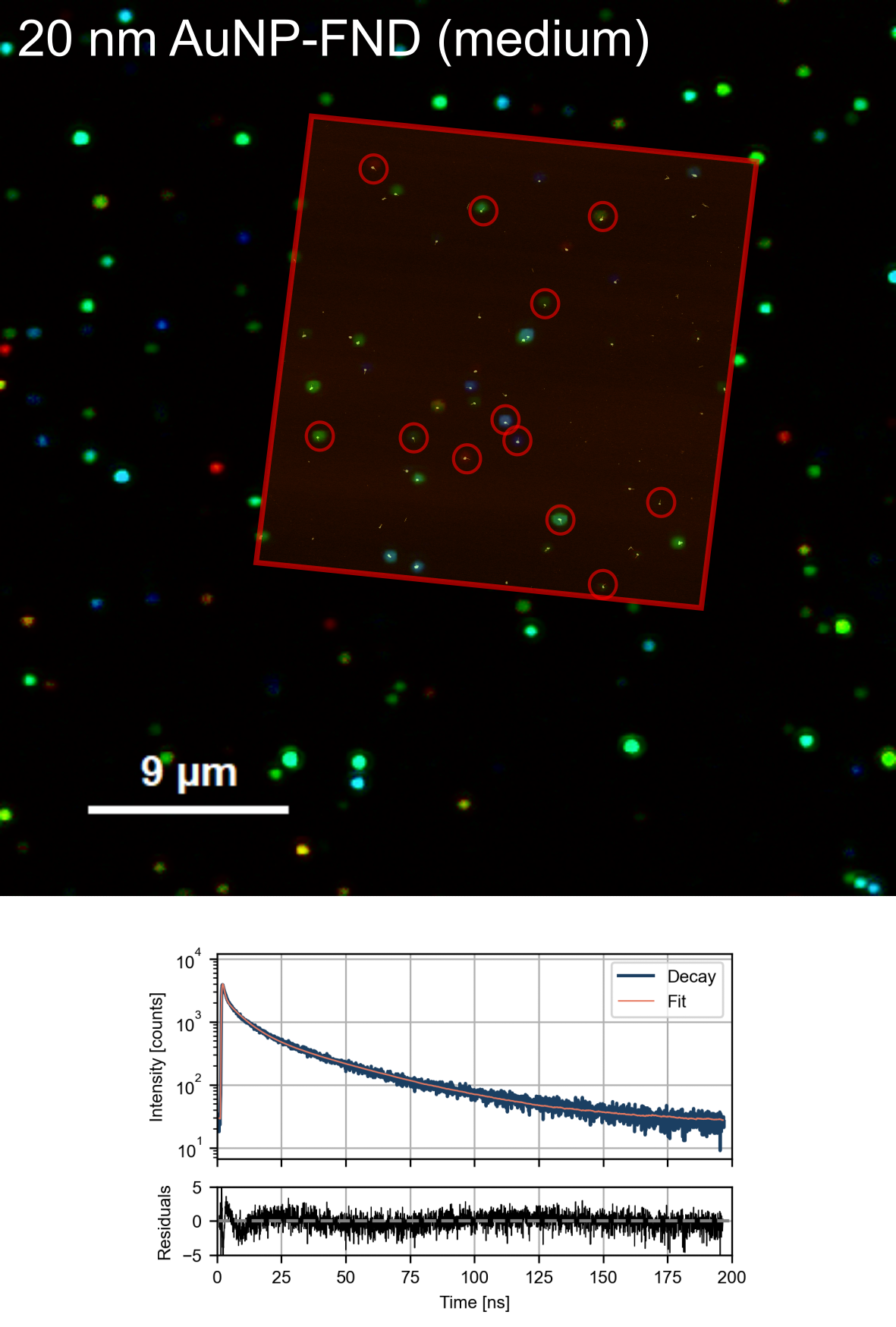}
    \caption{Overlay of high-resolution AFM image with FLIM measurement for assemblies consisting of 20 nm AuNP in the medium distance configuration (70 nm between binding sites). Red circles highlight selected structures. The graph shows the resulting fluorescence decay of the selected structures and the fit after analysis. Residuals represent goodness of fit with $\chi^2$ = 1.34}
    \label{fig:overlay_20nm-medium}
\end{figure}

\begin{figure}[ht]
    \centering
    \includegraphics[width=.8\linewidth]{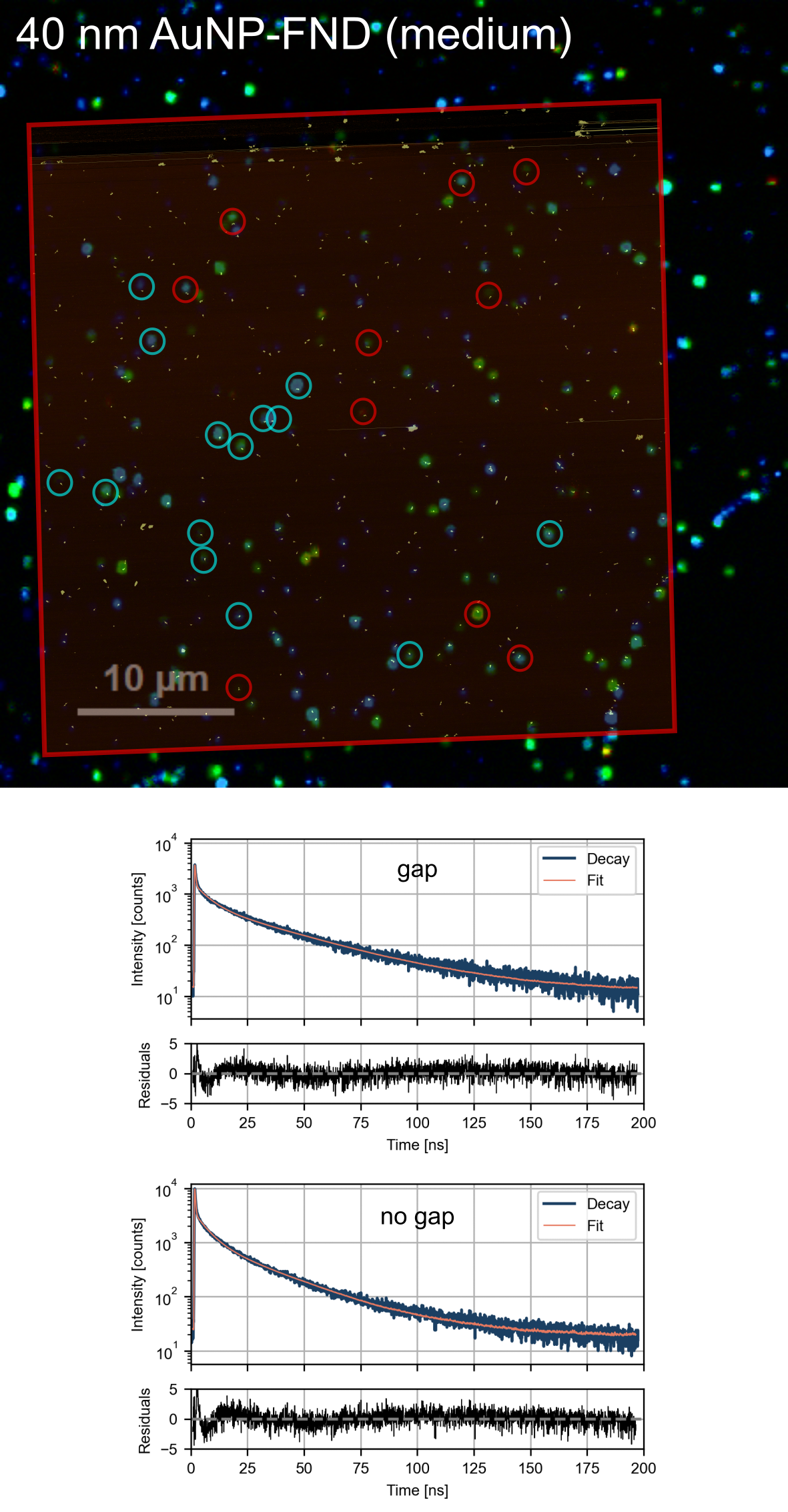}
    \caption{Overlay of high-resolution AFM image with FLIM measurement for assemblies consisting of 40 nm AuNP in the medium distance configuration (70 nm between binding sites). Red circles highlight selected structures with a visible gap between particles. Blue circles show structures without visible gap between particles. The graph shows the resulting fluorescence decay of the selected structures and the fit after analysis. Residuals represent goodness of fit with $\chi^2_{gap}$ = 1.21 and $\chi^2_{no\,gap}$ = 1.75}
    \label{fig:overlay_40nm-medium}
\end{figure}

\begin{figure}[ht]
    \centering
    \includegraphics[width=.8\linewidth]{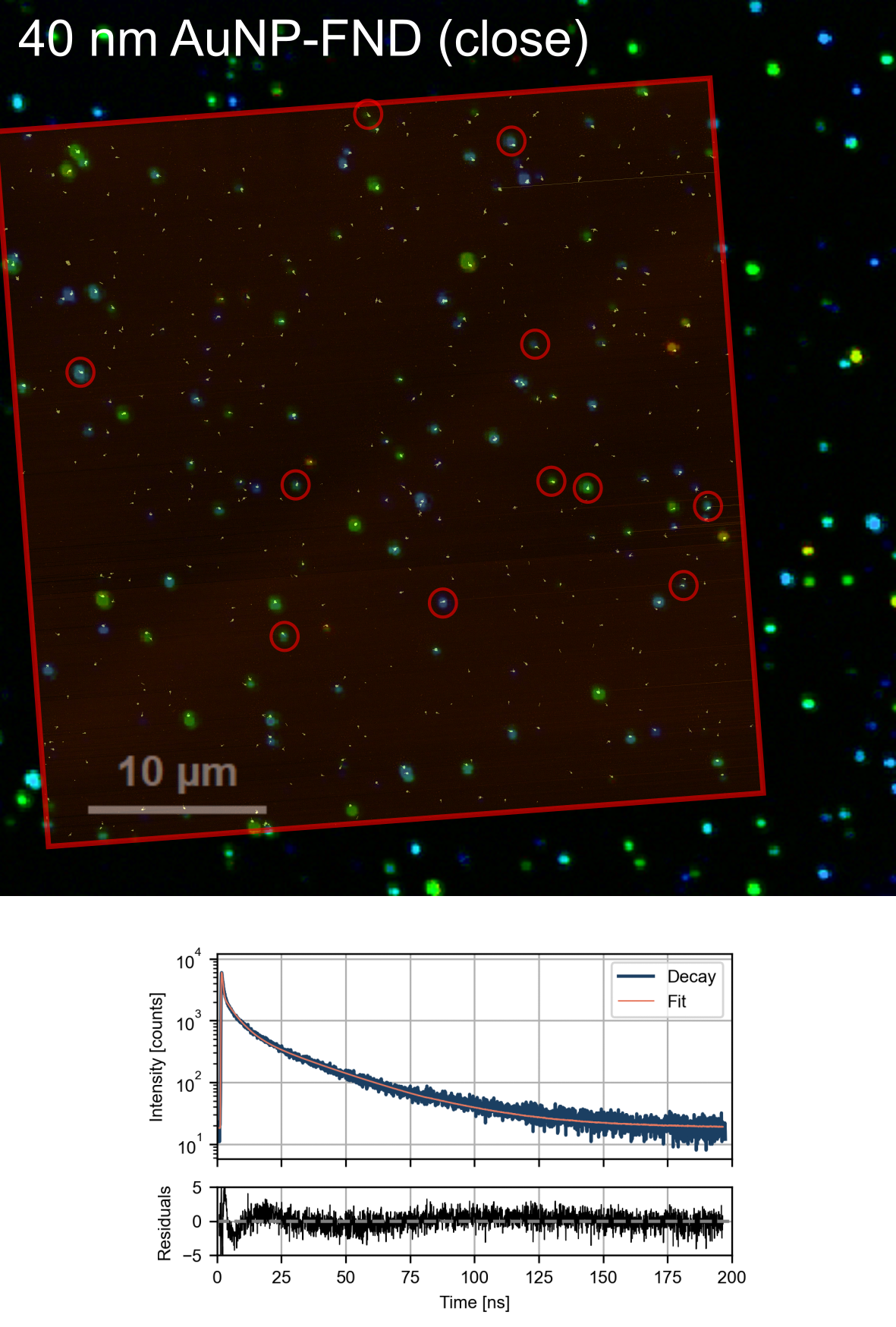}
    \caption{Overlay of high-resolution AFM image with FLIM measurement for assemblies consisting of 40\,nm AuNP in the close distance configuration (35\,nm between binding sites). Red circles highlight selected structures. The graph shows the resulting fluorescence decay of the selected structures and the fit after analysis. Residuals represent goodness of fit with $\chi^2$ = 1.43}
    \label{fig:overlay_40nm-close}
\end{figure}

\clearpage

\begin{figure}[ht]
    \centering
    \includegraphics[width=0.8\linewidth]{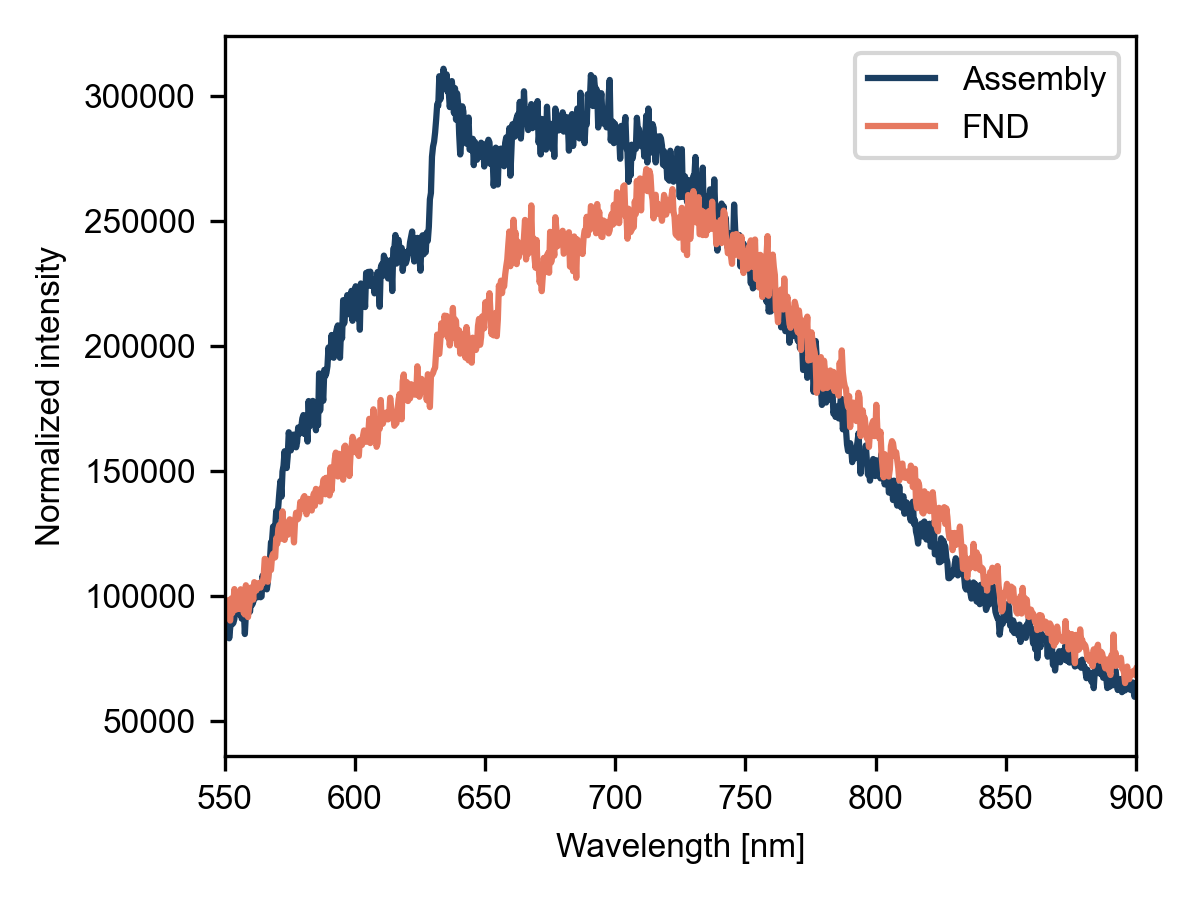}
    \caption{Single structure spectra for an assembly contining only one FND and an assembly containing a 40 nm AuNP and an FND in the "medium (no gap)" configuration.}
    \label{fig:spectra}
\end{figure}

\clearpage

\section*{Supplementary Note 5: 12-helix staple list and modifications}

\footnotesize
\begin{longtable}{lllc}
\caption{12-helix bundle staple list with nanoparticle binding staples indicated in the Modification column. An extended overview of the modified staple sequences can be found in Supplementary Table \ref{tab:SI_mod}} \\
\textbf{Start} & \textbf{End} & \textbf{Sequence}                                  & \textbf{Modification}  \endfirsthead 
\toprule
0[104]         & 0[85]        & CCTGGGGTGGAAACCATCGA                               &                        \\ \midrule
0[120]         & 11[130]      & AGCATAAAGACTCACATTAATGCTCACCGATTCAACCGATTG         & BS0 extended           \\ \midrule
0[146]         & 0[121]       & AGCCGAACAACAACATACGAGCCGGA                         &                        \\ \midrule
0[162]         & 11[172]      & AGAAAAGTAACCGAGGAAACGAGAGAGGATGTTTTCCCAGTC         & BS1 regular           \\ \midrule
0[188]         & 0[163]       & ATCGTAACCTACCGAAGCCCTTTTTA                         &                        \\ \midrule
0[204]         & 11[214]      & TGGTGTAGAGAGGGGACGACGGGAGCGCTAAAAAACAGGGAA         &                        \\ \midrule
0[232]         & 0[205]       & TTAGCGAACCTAATGGGATAGGTCACGT                       &                        \\ \midrule
0[246]         & 11[256]      & GAACGCGAGTTTTGAAGCCTTTCTCCTGTAATTTTTGTTAAA         &                        \\ \midrule
0[272]         & 0[247]       & TAGCTGATACTTATCCGGTATTCTAA                         &                        \\ \midrule
0[295]         & 1[286]       & ATCAATATGATATTCAGGTAGCT                            &                        \\ \midrule
0[316]         & 0[296]       & TCAACAGTAGGTCAAATCACC                              & BS2 shifted            \\ \midrule
0[330]         & 11[340]      & TATAAAGCCCCATATTTAACAGTATTTTTATACCAAAAACAT         &                        \\ \midrule
0[356]         & 0[331]       & ATATAACAGATACAAATTCTTACCAG                         &                        \\ \midrule
0[37]          & 11[45]       & TCAGGGCGATTTTTGGGGTCGACTTACATGGAAATAAATCCT         & BS0 extended           \\ \midrule
0[372]         & 11[382]      & CTGGAAGTTCGAACGAGTAGAATCGACCGTAGAAAACTTTTT         &                        \\ \midrule
0[398]         & 0[373]       & ATCAATATACAACTAAAGTACGGTGT                         &                        \\ \midrule
0[419]         & 11[424]      & TTTAATGGAAACAGTGCTTCTGTAAATTCTCCTTTGAGCTTCAAAGC    &                        \\ \midrule
0[440]         & 0[420]       & TTTTGCCAGTGAATTACCTTT                              &                        \\ \midrule
0[456]         & 11[466]      & GCTTTTGCAATGTTTAGACTGCAAATTATTTCGGGAGAAACA         &                        \\ \midrule
0[482]         & 0[457]       & TCATTTTGCAACCAAAATAGCGAGAG                         &                        \\ \midrule
0[498]         & 11[508]      & AGTTTGAGTCAGAAGGAGCGGTTTAATGCACTAACGGAACAA         &                        \\ \midrule
0[524]         & 0[499]       & ATAGGCTGGCTGACCTTCTTTTAAA                          &                        \\ \midrule
0[538]         & 11[552]      & TACAGACGATATTCATTACTGTTTGAGGAAGGTTATCTAAAA         &                        \\ \midrule
0[559]         & 0[539]       & TCAGAGGACAGATGAACGGTG                              & BS3 shifted            \\ \midrule
0[583]         & 11[591]      & GAGTTAAAGGAAAGACAGCATCCGCGAAATCTACGAAGGCAC         & BS3 regular            \\ \midrule
0[602]         & 4[590]       & GGTCGCTGCGCCGACTTAAACAAGAGGCTAGGTGAGGCGGT          &                        \\ \midrule
0[624]         & 0[603]       & TTTTCACGCATAACCGATATATTC                           &                        \\ \midrule
0[63]          & 0[38]        & CGACAGAATCGGCGAAAAACCGTCTA                         &                        \\ \midrule
0[84]          & 1[77]        & TAGCAGCACCGTAATTCAGACTG                            &                        \\ \midrule
1[134]         & 4[128]       & CCACAATGTGAAATCCAGTCTCGTGCCAGCTG                   &                        \\ \midrule
1[176]         & 4[170]       & ATCTGTCAAGAAAGAACTGGACTCCTTATTAC                   &                        \\ \midrule
1[218]         & 4[213]       & CCGTCCCTCCGTGGCACTCCCCGGCACCGCT                    &                        \\ \midrule
1[260]         & 4[254]       & AAGGAACCCAATAGCACCCATATCCTGAATCT                   &                        \\ \midrule
1[287]         & 11[301]      & ATTTTAACACTCATTAGAAACCAATCAAT                      & BS2 regular            \\ \midrule
1[302]         & 4[303]       & ACAGGCATTCAAAGGTCATTCTGGA                          &                        \\ \midrule
1[312]         & 11[322]      & TTGCGGGTATCTGTCTTTCCTTATC                          & BS2 regular            \\ \midrule
1[344]         & 4[340]       & CGTTTTAGAAAAAAGGCATTTAATAAGAGA                     &                        \\ \midrule
1[386]         & 4[380]       & TATGTGAATGCTGTTCGCAAACCTGTTTAGCT                   &                        \\ \midrule
1[424]         & 4[422]       & ATTTCATTAGACAAAATCTTAGAACATAGCGATAGC               &                        \\ \midrule
1[470]         & 4[464]       & TAAAGGCATAACCCGTCATAGAATCCCCCTCA                   &                        \\ \midrule
1[50]          & 4[44]        & AAAGAATATTAAAAACCCTACCGATTTAGAGC                   &                        \\ \midrule
1[513]         & 4[506]       & GTAATCTTTGCCTCAGATGATCAATATAATC                    &                        \\ \midrule
1[554]         & 4[552]       & TGAAGTCATAAGGTCATTCACTTGCCCT                       &                        \\ \midrule
1[78]          & 11[91]       & TAGCGAAAAAGAAGTTTGCCCCAGCAGG                       & BS0 regular/extended   \\ \midrule
1[8]           & 0[8]         & TTTTTCACCCAAATCAAGTTGGCCCACTACGTGAACCATTTT         &                        \\ \midrule
1[92]          & 4[92]        & CAATCCACCATTACATAGCCCGTTTG                         &                        \\ \midrule
10[621]        & 1[624]       & TTTTATACACTTAACAACCATCGCCTTTT                      &                        \\ \midrule
11[109]        & 0[105]       & GGTGGCGACACTGGGTTCCGAAATTGAGCTATGTAAAG             &                        \\ \midrule
11[131]        & 7[144]       & AGGGAGGGAGTGAATTCACGGAAACATATATATAACGTGCTT         & BS0 extended           \\ \midrule
11[15]         & 10[15]       & TTTTCTTGATATTCACAAACCTGTCAGACGATTGGCTTTT           &                        \\ \midrule
11[151]        & 0[147]       & GACGGCCAGGCCACGAAATTATTCGAAGGAAAGCAGAT             &                        \\ \midrule
11[173]        & 7[186]       & ACGACGTTGTGCAGGTTGCGGGCCGCAACTAGGAGGCCGATT         &                        \\ \midrule
11[193]        & 0[189]       & CCAGTAACATATAATGCCAAGCTTCCAGTTTTGGGCGC             &                        \\ \midrule
11[215]        & 7[228]       & GCGCATTAGTCAGAGGTCCCAATACAAAATGAATCCTGAGAA         &                        \\ \midrule
11[236]        & 0[233]       & ACTGATTAAGCCGCAACACCCTGCGGGAGGGCGTT                &                        \\ \midrule
11[257]        & 7[270]       & TCAGCTCATTTCGCGTAGCCCCAATGTACCAAAAGAGTCTGT         &                        \\ \midrule
11[277]        & 0[273]       & TAGGAGCATGCGCGAACGCCATCACGGAGAGACCGTTC             &                        \\ \midrule
11[302]        & 7[305]       & AATCGGTAAACCACAGAACGCAATAAACTTCT                   &                        \\ \midrule
11[323]        & 0[317]       & AGGTTGGAATCTTCCAAGAAAGAATCGAACGC                   & BS2 shifted            \\ \midrule
11[341]        & 7[354]       & TATGACCCTACGCAAGAATAAATACTAATACTCAAACTATCG         &                        \\ \midrule
11[361]        & 0[357]       & GCGGGACGCGGTAGAAGAAGCCTTAATTCTGTCATTCC             &                        \\ \midrule
11[383]        & 7[396]       & CAAATATATATGGTTTTAACCTCTCATAGGCCGCCAGCCATT         &                        \\ \midrule
11[403]        & 0[399]       & TCATCAATTCTGTTTTTCTGACCTATAACCTTACATAA             &                        \\ \midrule
11[425]        & 7[438]       & GAACCAGACAGTACCTTAGTCAGAATCAGGACATTTTGACGC         &                        \\ \midrule
11[445]        & 0[441]       & TCCAATTACACACTTCAGGTCAGGTAGTAAAAAAGAAG             &                        \\ \midrule
11[46]         & 7[60]        & CATTAAAGCCTCCAGTAAACCGCCCCTCCCTCGGTCACGCTGC        & BS0 extended           \\ \midrule
11[467]        & 7[480]       & ATAACGGATATCGCGCTAAAACACTACCATGCAGATTCACCA         &                        \\ \midrule
11[487]        & 0[483]       & CTTTGACGAAGAAAAAATACCAAGAAACCACAACATTA             &                        \\ \midrule
11[509]        & 7[522]       & CATTATTACATACCACCCTTATGTTTCAACTGGCCAACAGAG         &                        \\ \midrule
11[534]        & 0[525]       & TGAGGATAATCAGTTGAGAAGAACCGCAGGCGC                  &                        \\ \midrule
11[553]        & 7[557]       & TATCTTTCAATAGACTTGCTGGCAAATGATACG                  &                        \\ \midrule
11[574]        & 0[560]       & CTGCCACCGAATAATAGATTAAGCAGCGCCGCTTTTGCGGGA         & BS3 shifted            \\ \midrule
11[592]        & 7[602]       & CAACCTAAAACTGATAAGTACAACGGACTAAAATGCGCG            &                        \\ \midrule
11[71]         & 0[64]        & GCGCTGTACCATCTCTGAATTTTAGCGCAGTAG                  &                        \\ \midrule
11[92]         & 7[101]       & CGAAAAATCCCTTCACCAGTTGGGCGCCGCCGCTACAG             & BS0 extended           \\ \midrule
2[112]         & 2[113]       & ACCAGTAGCTGAGACTCCTCAAGAGAAGGCCTGCGTTGAATC         &                        \\ \midrule
2[154]         & 2[155]       & CATAGCTGCTCAGTACCAGGCGGATAAGTATCAATAATTGGT         &                        \\ \midrule
2[196]         & 2[197]       & GCCCAATAAGTATAGCCCGGAATAGGTGTACACAGTATTTAA         &                        \\ \midrule
2[280]         & 2[281]       & GGAATCATCTCAGAGCCACCACCCTCATTGATGAGAGACGTA         &                        \\ \midrule
2[35]          & 2[12]        & CCGAAGTTTTAACGGGGTCATTTT                           &                        \\ \midrule
2[364]         & 2[365]       & GGAATCATAACTACAACGCCTGTAGCATTACTTTAGTTCACC         &                        \\ \midrule
2[406]         & 2[407]       & TTAATTGCAGCGTAACGATCTAAAGTTTTATTCGTCGCGAGC         &                        \\ \midrule
2[448]         & 2[449]       & ACAAACATAAATGAATTTTCTGTATGGGAACGATAGCGTGAA         &                        \\ \midrule
2[490]         & 2[491]       & TAAGAGCAAGTTTCAGCGGAGTGAGAATATAAATTATCATAG         &                        \\ \midrule
2[574]         & 2[575]       & GCGCAGACGGCTCCAAAAGGAGCCTTTAAACGGAACGACGAG         &                        \\ \midrule
2[617]         & 11[621]      & TTTTGAGGTGAATTTCAATGACATCATCGCCGAAAGAGGCAAAAGATTTT &                        \\ \midrule
3[12]          & 2[36]        & TTTTGTGCCTTGAGTAACAGTGCCCGTATAGAGGTG               &                        \\ \midrule
3[137]         & 11[150]      & CGGGGTTTTGTTTCCTGGTTACCAATTAAAGAGGTAAATATT         & BS0 extended           \\ \midrule
3[179]         & 11[192]      & GGTTGATATAATAAGAGGCATCTGGCATGCCTAAAACGACGG         & BS1 regular           \\ \midrule
3[222]         & 11[235]      & AGGAGGTTTTCGGATTCGACTTGAACAAAGACGGGAGAATTA         &                        \\ \midrule
3[240]         & 3[239]       & ACCCTCAGAACCAAAAATCAAAGTAACAACCCGAGTACCGCC         &                        \\ \midrule
3[263]         & 11[276]      & AACCGCCACCTACCGCGTTAATGCAAAATAATTTTTAACCAA         &                        \\ \midrule
3[300]         & 1[311]       & GATAGCAAGCCCAATGGTAAAGTTAA                         &                        \\ \midrule
3[321]         & 3[320]       & CCATGTACCGTAACATAACGCCAAGAGTAATGTGTAAGGAAC         & BS2 shifted            \\ \midrule
3[347]         & 11[360]      & CACCAGTACAAATTACTGATTCCCTATTTCAGTAATACTTTT         &                        \\ \midrule
3[389]         & 11[402]      & CCTCATAGTTTGAATATTGAGTGAAAATTTATTTAGTTAATT         &                        \\ \midrule
3[431]         & 11[444]      & AGACGTTAGTCAAGAAAGGGGTAAATTAGAGCGGAAGCAAAC         &                        \\ \midrule
3[473]         & 11[486]      & AACTTTCAACACACTATAACAAAGTTACAAATCGCCTGATTG         &                        \\ \midrule
3[520]         & 11[533]      & GGAATTTAAATCCTTGACATTTAGGAAGGTAGAAAGATTCAT         & BS3 regular            \\ \midrule
3[532]         & 3[531]       & AATAATTTTTTCACACCCAAATCCGACAACTCGTATGCGAAT         & BS3 shifted            \\ \midrule
3[557]         & 11[573]      & CCAAAAAAAAGGTCAATCACCCTCGAGCCGTAGGAGCACTAACAA      &                        \\ \midrule
3[57]          & 11[70]       & CTGCCTAGTCCACGTTTGCCTTACCGTAGAATGGAAAGCGCA         & BS0 regular/extended   \\ \midrule
3[598]         & 3[617]       & TTATCAGCTTGCTTTCTTTT                               &                        \\ \midrule
3[69]          & 3[68]        & GAACCTATTATTCTGCGGCGTTTTGTTTGGAACAAGATTTCG         &                        \\ \midrule
3[91]          & 11[108]      & AAAGTATTAAGAGGCACCATTTAATGAGCGGCAAATCCTGTTTGAT     & BS0 extended           \\ \midrule
4[112]         & 4[113]       & CAACGCGCGGATTAGCCCTTGGAGAGGCCAGCAACGATCGGC         &                        \\ \midrule
4[127]         & 7[123]       & CATTAATGACTCACTGCGTTTGGGAATTAGATGGTTGCTTTG         & BS0 extended           \\ \midrule
4[151]         & 4[152]       & TAGAAAAACCTGGGGAATACATACGTAATCAAATAGCAAACG         &                        \\ \midrule
4[169]         & 7[165]       & GCAGTATGTCGGAATAAGCCGAGCTCGAATGAATCAGAGCGG         & BS0 extended           \\ \midrule
4[196]         & 4[197]       & AGGCAAAGCGTTAAGCATGACCATTCGAATTGAGCGGAAACC         &                        \\ \midrule
4[212]         & 7[207]       & TCTGGTGCCGGCCTCAGCAAGATAACCCACATTAGACAGGAAC        &                        \\ \midrule
4[235]         & 4[236]       & TCTTTCCGCTTTAGCCAAGAGCCTGTGAGCGGATAACGAGCG         &                        \\ \midrule
4[253]         & 7[249]       & TACCAACGCTTAGTTGATCATCAACATTAATAATCAGTGAGG         &                        \\ \midrule
4[280]         & 4[281]       & GGTAATCGTAAATTTGCTACAAACTAGTTTTCATTCATGAAC         &                        \\ \midrule
4[302]         & 7[290]       & GCAAACAAGAGAATCGTACAAAGTTGCAAGCCGTTTTAAATTAACCGT   &                        \\ \midrule
4[323]         & 4[324]       & AAAGGTAAAGTAGAGTGCCTGAATTCTGAATGCCTCACGACA         &                        \\ \midrule
4[339]         & 7[326]       & ATATAAAGTACTGTAATTCATATATTTTAAATTCACT              &                        \\ \midrule
4[364]         & 4[365]       & GCGCGAGCTGGCCAGTTCGAAAAAGGTGAATAAATGTTTGGG         &                        \\ \midrule
4[379]         & 7[375]       & ATATTTTCAACCATTAGAAAGGCGTTAAATGTAATATCCAGA         &                        \\ \midrule
4[406]         & 4[407]       & GAGAAGAGTCCAATAATGGTAATAGTGTGGCTTATAGACGCT         &                        \\ \midrule
4[421]         & 7[417]       & TTAGATTAATTAATTACCGTCATTTTTGCGAAAACGCTCATG         &                        \\ \midrule
4[43]          & 7[39]        & TTGACGGGGAAGCACTGCATACAGGAGTGTCTAGGGCGCTGG         & BS0 extended           \\ \midrule
4[448]         & 4[449]       & TCAGAAAACGGAAAATCCTTAGAATGAAAGATGATCAACAGT         &                        \\ \midrule
4[463]         & 7[459]       & AATGCTTTACAATACTTTTACCTGAGCAAATGAAATGGATTA         &                        \\ \midrule
4[490]         & 4[491]       & ATACTTCTGATCATTAATATATAATGGACGAGGCATTGGATT         &                        \\ \midrule
4[505]         & 7[501]       & CTGATTGTTCATATTCATCGCCAAAAGGAACCAGTAATAAAA         &                        \\ \midrule
4[524]         & 4[525]       & ATAATTCATGGCTGGGCTTAACAATTAAAGAACGAGTAGTAA         &                        \\ \midrule
4[551]         & 7[536]       & GACGAGAAACACCCGTAACATATATTAGACTTTATTCTG            &                        \\ \midrule
4[574]         & 4[575]       & CTGCAACAGTTAAGGGTGAAGCCACGCGCCGGAAGGCACCGC         &                        \\ \midrule
4[589]         & 7[585]       & CAGTATTAAGTAGCAATTCTCCATGTTACTTTTGAATGGCTA         &                        \\ \midrule
4[621]         & 5[621]       & TTTTGCAGAAGATAAAACAGTTCGAACGAACCACCATTTT           &                        \\ \midrule
4[71]          & 4[72]        & TCACCGGAACCAGCCCAAGGGAGAGCCATGTTCCACACAAAA         &                        \\ \midrule
4[91]          & 7[75]        & CCATCTTTTCATAATTCGGCATGTATAGGGTTGAGTCCACAC         & BS0 extended           \\ \midrule
5[15]          & 4[15]        & TTTTGCGAGAATGGTAATTAAAAGCCGGCGAACGTGTTTT           &                        \\ \midrule
6[311]         & 8[300]       & ACGACGCCTGTTAGATAAGTCCTGAA                         & BS2 regular            \\ \midrule
6[566]         & 8[552]       & GCCAGCAAACCTCACCTCAATCAATATCT                      & BS3 regular            \\ \midrule
6[624]         & 9[628]       & TTTTATCGCCATTAAAAATACGAGGAGATTTGAAAACACTCATCTTTTTT &                        \\ \midrule
7[102]         & 8[89]        & GGCGCGTACTAGCGGTTTGCGTATGAGACGGATTGCCCTTCACCGCC    & BS0 regular/extended   \\ \midrule
7[124]         & 3[136]       & ACGAGCACGAAAGAAACCCGCTTTTGTTATATTAGGATTAG          &                        \\ \midrule
7[145]         & 8[135]       & TCCTCGTTATCATAAAGGTGGCATAAGTTTAATCAATAGAAA         & BS0 extended           \\ \midrule
7[166]         & 3[178]       & GAGCTAAACGTTGGGACCCAAAACAATGAAGCCGTCGAGAG          &                        \\ \midrule
7[187]         & 8[177]       & AAAGGGATTAGCCATTCAGGCTGCTCTTCGAGCTGGCGAAAG         & BS1 regular           \\ \midrule
7[208]         & 3[221]       & GGTACGCCAAAACAGCGAAGATCGGAACAAATCACCGTACTC         &                        \\ \midrule
7[229]         & 8[219]       & GTGTTTTTAATAATTTGCCAGTTCCAAATATTTTGTTTAACG         &                        \\ \midrule
7[250]         & 3[262]       & CCACCGAGTCCGGTTGCTATTTTGCAAGCAGCCACCCTCAG          &                        \\ \midrule
7[271]         & 8[261]       & CCATCACGCTACATGTCAATCATAAAACAGTAAGCAAATATT         &                        \\ \midrule
7[291]         & 3[299]       & TGTAGCAATACAACATGGCTATCAAGGGTGATTCAGG              & BS2 shifted            \\ \midrule
7[306]         & 6[312]       & TTGATTAGTAATAACAGCTCCAGACG                         &                        \\ \midrule
7[327]         & 3[346]       & TGCCTGAGTAGAAGAAGTAGTAGTAGGCAGGCCTGTTCTGAGTTTCGT   &                        \\ \midrule
7[355]         & 8[345]       & GCCTTGCTGAAGGCATCAATTCTCATACAGGAATTAGCAAAA         &                        \\ \midrule
7[376]         & 3[388]       & ACAATATTATCTGAGAGATACATTAGCTCACCACAGACAGC          &                        \\ \midrule
7[397]         & 8[387]       & GCAACAGGAGAAATTTATCAAAACGGCTTATATAACTATATG         &                        \\ \midrule
7[40]          & 3[56]        & CAAGTGTAGCAGAGCCAAATCGGGAACGTGAACAGTTAATGCCCC      &                        \\ \midrule
7[418]         & 3[430]       & GAAATACCTTCTTTACATTTTCCTAATTACGTCGTCTTTCC          &                        \\ \midrule
7[439]         & 8[429]       & TCAATCGTCAGCCATAAATCAAAAAGCAAATCAAAAAGATTA         &                        \\ \midrule
7[460]         & 3[472]       & TTTACATTGATCAAAAGCGGAATCTCGTTTTTTTGCTAAAC          &                        \\ \midrule
7[481]         & 8[471]       & GTCACACGATTAAGGGTTAGAACGAAATAATAGATTTTCAGG         &                        \\ \midrule
7[502]         & 3[519]       & GGGACATTCTTTAATCCTGATTACGAACGTGAAAGGAACAACTAAA     &                        \\ \midrule
7[523]         & 8[513]       & ATAGAACCCCAGAGATGGTTTAACGATTTTTCATTATACCAG         &                        \\ \midrule
7[537]         & 3[556]       & ACCTGAAAGCGTAAGAAAAAATCAAGCTGCGAACCGAGTTGAAAATCT   & BS3 shifted            \\ \midrule
7[558]         & 6[567]       & TGGCACAGACAATATTTATGAGA                            &                        \\ \midrule
7[586]         & 3[597]       & TTAGTCTTTAGACTTTCGGCTACGCTTGATTTGTATCGGT           &                        \\ \midrule
7[603]         & 7[624]       & AACTGATAGCCCTAAAACTTTT                             &                        \\ \midrule
7[61]          & 8[51]        & GCGTAACCAGTCCACCGGAACCGACCCTCACCTCAGAGCCGC         & BS0 extended           \\ \midrule
7[76]          & 3[90]        & CCGCCGCGCTTAATGCAGGGTGTTTCGGTGCAAGGCAAACATG        &                        \\ \midrule
7[8]           & 6[8]         & TTTTGCGAAAGGAGCGGGCGACAGGAAGGGAAGAAATTTT           &                        \\ \midrule
8[113]         & 8[114]       & CAAAAGCTGGCAACGACAAAATGAGCCACAGTTTACCAGCGC         & BS0 extended           \\ \midrule
8[134]         & 1[133]       & ATTCATATGAAGACACATCACCGTCACAATT                    &                        \\ \midrule
8[155]         & 8[156]       & TAAGGTCACATTTTTTGGGTACCGGGTAGGCTGCAAGGCGAT         & BS0 extended           \\ \midrule
8[176]         & 1[175]       & GGGGATGTGCGATCGGCGACTCTCAATAGCT                    &                        \\ \midrule
8[197]         & 8[198]       & TTTAACGCCCTATTCAGAGAGATCAGAGTAAAAATAGCAGCC         & BS1 regular           \\ \midrule
8[218]         & 1[217]       & TCAAAAATGTTATTTAGTAATTGCGGATTGA                    &                        \\ \midrule
8[239]         & 8[240]       & TTGTCGATTAGAAATAAAATTCAGCTTTAAAACGTTAATATT         &                        \\ \midrule
8[260]         & 1[259]       & TAAATTGTATCAGAAACTGGCCTAGATATAG                    &                        \\ \midrule
8[281]         & 8[282]       & ATCCTTGTAGAAGATAATTTAAGAACAACAAAATAATATCCC         &                        \\ \midrule
8[299]         & 1[301]       & CAAGAAGCTAATGAGTACCGGGCCGGAG                       &                        \\ \midrule
8[320]         & 8[321]       & AAACAATTATCAAGCTAAACCCTCATTAAAGCCTCAGAGCAT         & BS2 regular            \\ \midrule
8[33]          & 8[12]        & GAGCCGCCGCCAGCATTGTTTT                             &                        \\ \midrule
8[344]         & 1[343]       & TTAAGCAATAACATCCGATAAAAATCATATG                    &                        \\ \midrule
8[365]         & 8[366]       & TCGCGCAAAGCAAGAAGACAAGATAAATCTATGCAAATCCAA         &                        \\ \midrule
8[386]         & 1[385]       & TAAATGCTGACCTTTTGAAATACGTTTTAAA                    &                        \\ \midrule
8[407]         & 8[408]       & AAATGGTTAGGTTGATCGCGTATAAGAGTGCCGAAAGACTTC         &                        \\ \midrule
8[428]         & 1[423]       & AGAGGAAGCACTATTATTAATTGAACA                        &                        \\ \midrule
8[449]         & 8[450]       & AGTATTGCAGCGGAACAGTACTTTCAATATAGATGAATATAC         &                        \\ \midrule
8[470]         & 1[469]       & TTTAACGTCTTGCACGAGAGGCGGACGACGA                    &                        \\ \midrule
8[491]         & 8[492]       & TCTATTGCGAGAAACGTTAATTACATAATGTGGGAAGAAAAA         &                        \\ \midrule
8[50]          & 1[49]        & CACCAGAACCCCTCAGAGCGTCACCAACGTC                    &                        \\ \midrule
8[512]         & 1[512]       & TCAGGACGTTGAATTAATTCAACAAATCAAGA                   &                        \\ \midrule
8[539]         & 8[540]       & AATCAACAGTCTGGCAAGAATGAAAGGTTAGAAGAATTGGCA         & BS3 regular            \\ \midrule
8[551]         & 1[553]       & GGTCAGGCATCACTAATACAACCAACTT                       &                        \\ \midrule
8[575]         & 8[576]       & GTAACAAACAATATAATACGTCGACCTGCATCCATTAAACGG         &                        \\ \midrule
8[597]         & 0[584]       & AAACAAAGTTTGAGGAAATTGTGTATAGTTGAGGCTTGCAGG         &                        \\ \midrule
8[628]         & 8[598]       & TTTTTGACCCCCAGCGATTATACCAAGCGCG                    &                        \\ \midrule
8[68]          & 8[69]        & CACCACGAGCCAAGCGGTGCCCGAGTTCTGAGAGAGTTGCAG         & BS0 regular/extended   \\ \midrule
8[88]          & 1[91]        & TGGCCTTCTTTTATAAATCAACGTCAC                        &                        \\ \midrule
9[12]          & 8[34]        & TTTTACAGGAGGTTGAGGCAGTTTGATGCACACCACCA             & BS0 extended           \\ \midrule
\bottomrule
\end{longtable}


\clearpage

\begin{figure}[ht]
    \centering
    \includegraphics[width=0.5\linewidth]{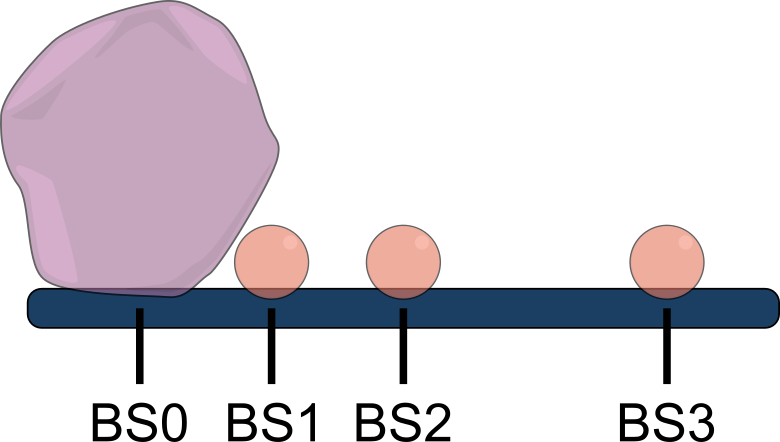}
    \caption{Sketch of binding site position along 12HB structure with most common nanoparticle species at their respective position.}
    \label{fig:SI_BS}
\end{figure}

\begin{table}[ht]
\caption{Staple extensions for binding AuNP and FNDs at different binding sites as shown in Supplementary Figure \ref{fig:SI_BS} for different configurations. }

\centering
\footnotesize
\begin{tabular}{cccc}
\textbf{Binding site}         & \textbf{Configuration} & \textbf{AuNP}                                & \textbf{FND}                                   \\ 
\toprule
\multirow{2}{*}{\textbf{BS0}} & regular                & AAA AAA AAA AAA AAA AAA AAA                  & \multirow{2}{*}{AAG~AAG~AAG~AAG~AAG~AAG~AAG~}  \\
\cmidrule(lr){2-2}
                              & extended               & -                                            &                                                \\
\midrule
\textbf{BS1}                  & regular                & AAA AAA AAA AAA AAA AAA AAA                  & -                                              \\
\midrule
\multirow{2}{*}{\textbf{BS2}} & regular                & \multirow{2}{*}{AAA AAA AAA AAA AAA AAA AAA} & \multirow{2}{*}{-}                             \\
\cmidrule(lr){2-2}
                              & shifted                &                                              &                                                \\
\midrule
\multirow{2}{*}{\textbf{BS3}} & regular                & \multirow{2}{*}{AAA AAA AAA AAA AAA AAA AAA} & \multirow{2}{*}{-}                             \\
\cmidrule(lr){2-2}
                              & shifted                &                                              &                                                \\
\bottomrule
\end{tabular}

\label{tab:SI_mod}
\end{table}